\documentstyle[12pt,axodraw,psfig]{article}
\begin{document}
\baselineskip 0.6cm
\newcommand{\gsim}{ \mathop{}_{\textstyle \sim}^{\textstyle >} }
\newcommand{\lsim}{ \mathop{}_{\textstyle \sim}^{\textstyle <} }
\newcommand{\vev}[1]{ \left\langle {#1} \right\rangle }
\newcommand{\bra}[1]{ \langle {#1} | }
\newcommand{\ket}[1]{ | {#1} \rangle }
\newcommand{\EV}{ {\rm eV} }
\newcommand{\KEV}{ {\rm keV} }
\newcommand{\MEV}{ {\rm MeV} }
\newcommand{\GEV}{ {\rm GeV} }
\newcommand{\TEV}{ {\rm TeV} }


\begin{titlepage}

\begin{flushright}
UCB-PTH-01/04 \\
LBNL-47413 \\
HUTP-01/A006 \\
UW/PT-01/01 \\
\end{flushright}

\vskip 0.4cm

\begin{center}
{\Large \bf  Finite Radiative Electroweak Symmetry Breaking \\
  from the Bulk}

\vskip 1.0cm

\def\thefootnote{\fnsymbol{footnote}}
{\bf 
Nima Arkani-Hamed$^{a,b,c}$, 
Lawrence Hall$^{a,b}$, 
Yasunori Nomura$^{a,b}$\footnote{Research Fellow,
  Miller Institute for Basic Research in Science.}, \\
David Smith$^{a,b}$ and 
Neal Weiner$^{d}$
}

\vskip 0.5cm

$^a$ {\it Department of Physics, University of California,
                 Berkeley, CA 94720}\\
$^b$ {\it Theoretical Physics Group, Lawrence Berkeley National Laboratory,
                 Berkeley, CA 94720}\\
$^c$ {\it Jefferson Physical Laboratory, Harvard University, 
                 Cambridge, MA 02138}\\
$^d$ {\it Department of Physics, University of Washington, 
                 Seattle, WA 98195}\\

\vskip 1.0cm

\abstract{A new physical origin for electroweak symmetry breaking is
proposed, involving compact spatial dimensions of scale $1/R \approx 1$
TeV. The higher dimensional theory is supersymmetric, and hence
requires the top-quark Yukawa coupling to be localized on some
``Yukawa brane'' in the bulk. The short distance divergence in the
Higgs-boson mass is regulated because supersymmetry is unbroken in the
vicinity of this Yukawa brane. A {\em finite}, negative Higgs
mass-squared is generated radiatively by the top-quark supermultiplet
propagating a distance of order $R$ from the Yukawa brane to probe
supersymmetry breaking. The physics of electroweak symmetry
breaking is therefore closely related to this top propagation across
the bulk, and is dominated by the mass scale $1/R$, with exponential
insensitivity to higher energy scales. The masses of the superpartners
and the Kaluza-Klein resonances are also set by the mass scale $1/R$,
which is naturally larger than the $W$ boson mass by a loop factor.

Explicit models are constructed which are highly constrained and
predictive. The finite radiative correction to the Higgs mass is
computed, and the Higgs sector briefly explored.
The superpartner and Kaluza-Klein resonance spectra are
calculated, and the problem of flavor violation from squark and
slepton exchange is solved. Important collider signatures include 
highly ionizing charged tracks from stable top squarks, and events 
with two Higgs bosons and missing transverse energy.}

\end{center}
\end{titlepage}

\def\thefootnote{\arabic{footnote}}
\setcounter{footnote}{0}


\section{Introduction}

The physical origin of the mass scale of electroweak symmetry breaking
(EWSB) is unknown. 
In technicolor theories, the mass scale of EWSB has a direct
physical interpretation as the scale at which some new gauge force
becomes strong. In supersymmetric theories, the scale of EWSB is
related to the scale of supersymmetry breaking, although the
connection is often indirect and model dependent \cite{gravity, gauge}.
In this paper we introduce a mechanism which requires the EWSB scale 
to be directly connected to the length scale of a new compactified 
spatial dimension.

The physical mechanism for EWSB is also unknown. Precision electroweak 
data suggest that there is a perturbatively coupled Higgs boson 
\cite{Higgs-LEP}, and it is this possibility that we explore. 
The large value of the top-quark mass implies that the Higgs
boson has a Yukawa coupling to the top quark, $y_t$, which is close to
unity.  This leads to a one-loop quadratic divergence in the Higgs-boson
mass-squared parameter
\begin{equation}
  m_H^2 =  m_{H0}^2 - {N_c\, y_t^2 \over 4 \pi^2} \Lambda^2,
\label{eq:smhiggsmass}
\end{equation}
where the standard model is viewed as a low energy effective theory
valid up to some cutoff $\Lambda$, and $m_{H0}^2$ is the tree-level 
mass parameter. Given the negative sign of this
radiative correction, it is tempting to infer that it is this
radiative correction which is triggering EWSB. However, this
conclusion cannot be drawn --- the quadratic divergence implies that
the scale of the physics triggering EWSB is at $\Lambda$, where the
low energy effective theory breaks down and is unreliable. In this
sense, the standard model does not provide a theory of EWSB. For
$\Lambda$ larger than a few TeV, the cancellation between the tree
and radiative terms in Eq.~(\ref{eq:smhiggsmass}) becomes problematic.

In supersymmetric theories the quadratic divergence of the Higgs mass
parameter from the top-quark loop is cancelled by that from the top-squark 
loop.  The residual divergence is logarithmic
\begin{equation}
  m_H^2 =  m_H^2(\Lambda) - {N_c\, y_t^2 \over 4 \pi^2}\, 
    m_{\tilde{t}}^2 \ln {\Lambda^2 \over  m_{\tilde{t}}^2},
\label{eq:susyhiggsmass}
\end{equation}
with the mass scale of the radiative correction determined by the
top-squark mass, $m_{\tilde{t}}$ \cite{radewsb}. 
Here $\Lambda$ is again understood to be the ultraviolet cutoff 
of the supersymmetric theory containing soft supersymmetry-breaking 
parameters. Such theories have gained much attention 
over the last two decades because, for large values of the
logarithm, the radiative triggering of EWSB is reliably computed in the
low energy theory --- one has a theory of EWSB.

What is the energy scale of the physics that triggers EWSB in
supersymmetric theories? The logarithmic divergence implies that the
Higgs mass parameter can be viewed as running with scale, so that the 
physics of EWSB is the evolution of this parameter with energy. 
The energy region where most of this evolution occurs is model dependent 
--- it can be anywhere between the top-squark mass, which sets the mass 
scale of the electroweak vacuum expectation value, and the ultraviolet 
cutoff, $\Lambda$. 

In supersymmetric theories the quartic Higgs coupling is predicted, 
leading to a well known upper bound on the Higgs mass of about 135 GeV
\cite{Higgs-bound}.
Also, in the absence of fine-tuning, the physical Higgs mass-squared is
given by the size of the radiative correction, and is therefore
expected to be $(m_{\tilde{t}}^2/10) \ln (\Lambda^2 / m_{\tilde{t}}^2)$.  
For large values of the logarithm, as in gravity mediated theories, 
the top squark is not expected to be much heavier than the Higgs boson, 
so that typically $m_{\tilde{t}} \lsim 200$ GeV, conflicting with data. 
For small values of the logarithm, as in certain gauge mediated 
theories, it is most natural for the top squark to be a factor 3 
heavier than the Higgs boson. But in these theories the charged
slepton and winos are significantly lighter than the top squark, and
the direct searches at LEP imply that the top-squark mass is well above 
three times the Higgs mass. Thus, in the most-studied supersymmetric
theories, we already know that EWSB does not occur in the most natural
region of parameter space \cite{Barbieri:2000gf}. The amount of 
parameter tuning is modest, and this analysis fuels an expectation 
that superpartners may well be discovered soon.

In this paper we introduce a new mechanism for radiative EWSB in
supersymmetric theories. The divergence in the Higgs mass is cut off
by the scale of a new compact dimension of TeV size. The
radiative Higgs mass is calculable and finite, and dominated by physics
at the TeV scale. Our mechanism relies on a departure from previous
models with TeV-sized extra dimensions \cite{Antoniadis:1990ew, 
TeV-1, TeV-2, TeV-pheno}, which have brane-localized matter fields. 
Two crucial ingredients are required in our framework:
\begin{itemize}
\item The virtual top quarks and top squarks in the radiative diagrams 
for the Higgs-boson mass propagate in the new compact dimension.
\item The top-quark Yukawa coupling and the breaking of supersymmetry
are not located at the same point in the bulk.
\end{itemize}
This implies that, in the radiative Higgs mass calculation, 
the virtual propagators of the top-quark multiplet
must sample the bulk far from the Yukawa interaction to avoid an exact
supersymmetric cancellation. If this distance scale is $D$, then the
contributions from large virtual 4-momentum, $k$, to the Higgs mass
are exponentially suppressed by $e^{-kD}$. 
The resulting radiative
contribution to the Higgs mass-squared parameter is found to be 
{\it finite}: $m_H^2 =  m_{H0}^2 - C (N_c y_t^2 / \pi^2) (1 / D)^2$,
where $C$ is a model dependent parameter of order unity.
Comparing with the standard model result of Eq.~(\ref{eq:smhiggsmass}),
we see that the quadratic divergence is regulated by the spatial 
separation in the bulk.  
Contrary to previous models, the quadratic divergence of the Higgs 
boson mass is directly cut off at the distance scale $D$; in no energy 
region is the Higgs boson mass logarithmically sensitive to the cutoff 
of the effective theory.

What is the scale $D$ which governs the separation of supersymmetry
breaking and flavor breaking? 
In this paper we take the bulk to preserve supersymmetry, forcing the
top-quark Yukawa coupling to be localized on some ``Yukawa brane''.
One possibility is that $D$ is simply the distance across the
bulk from the Yukawa brane to the closest brane on which there is 
supersymmetry breaking. However, we have an alternative picture in mind. 
The relevant scale is the distance scale on which the top-quark multiplet
feels supersymmetry breaking. We assume that the top multiplet feels
supersymmetry breaking via the form of its mode expansion in the
bulk --- ie supersymmetry breaking forces the Kaluza-Klein (KK) expansion 
of the top quarks to differ from that of the top squarks. Since the
dimensionful parameter of the KK mode expansion is the radius of the
bulk, $R$, we expect $D \approx R$. In this paper we study theories
with a one dimensional bulk, taken to be the $S^1/Z_2$
orbifold, so that the distance is the length of the orbifold, 
$D = \pi R$, giving
\begin{equation}
  m_H^2 = - C  {N_c\, y_t^2 \over \pi^4} \left( { 1 \over R} \right)^2.
\label{eq:kkhiggsmass}
\end{equation}
We have set $m_{H0}^2 = 0$ --- our EWSB mechanism only works if the 
tree-level Higgs soft mass is small, and, in the theories considered 
in this paper, it vanishes.\footnote{This mechanism for EWSB has been 
used in the context of an $S^1/(Z_2 \times Z_2')$ orbifold 
\cite{Barbieri:2000vh}.}

Electroweak symmetry is broken radiatively by the large top-quark
Yukawa interaction, but locality exponentially cuts off supersymmetry
breaking at short distances. The top multiplet propagators are
supersymmetric at high energies, apart from $e^{-k \pi R}$
corrections, so that EWSB is necessarily broken by physics {\it at the 
compactification scale.} From Eq.~(\ref{eq:kkhiggsmass}) we see that 
there is a very close relationship between the Higgs mass and the 
compactification scale: the physical Higgs mass is about $0.2(1/R)$. 
Our EWSB mechanism has no naturalness problem; the Higgs boson is 
lighter than the KK resonances by a loop factor.

What is the general structure of the theory just above the weak scale?
This is the crucial question for future collider physics. With 
conventional radiative EWSB there are superpartners, since we have an 
energy region described by a 4d supersymmetric theory. However, with 
our mechanism there is no energy interval where physics is described 
by a 4d supersymmetric theory. Just above the weak scale we have a 
5d theory, which has two supersymmetries from the 4d viewpoint.
As well as the usual superpartners, there are the ``$N=2$'' partners 
and KK resonances, all having mass splittings determined by $1/R$.
The spectrum of this large number of states is model dependent, 
but, in the models described below, is given in terms of just a few free 
parameters. The presence of these extra states is a necessary consequence 
of our new EWSB scheme, with the Higgs mass divergence regulated by 
5d supersymmetry, broken at the compactification scale $1/R$. The 
physics of EWSB is the physics of the spectrum of these states near 
$1/R$. Unlike the 4d supersymmetric case, physics at scales much larger 
than the weak scale is irrelevant.

In section \ref{section:rad-cor} we perform a calculation of the Higgs 
mass-squared using KK towers of the top quarks and top squarks which 
are shifted relative to each other to reflect supersymmetry breaking. 
We show in detail how the Bose and Fermi loop contributions, when 
summed over the entire KK tower, lead to the finite Higgs mass result 
of Eq.~(\ref{eq:kkhiggsmass}), with a parameter $C$ close to unity. 
This provides a general illustration of our EWSB mechanism, but leaves 
two issues open: what is the underlying mechanism of supersymmetry 
breaking, and what provides the restoring potential for the Higgs field?

We study two possibilities for supersymmetry breaking --- local and 
non-local in the bulk.
In sections \ref{section:localized} and \ref{section:Z2R-SS} we study 
two explicit models, illustrating local and non-local supersymmetry 
breaking respectively. 
In section \ref{section:localized} supersymmetry breaking is localized 
on a three brane, and is coupled directly to the zero-mode top squark, 
resulting in a non-uniform profile for its wavefunction 
in the bulk. In section \ref{section:Z2R-SS} supersymmetry is broken
by the Scherk-Schwarz mechanism \cite{Scherk-Schwarz} 
using $R$ parity: under a translation about the circle by $2\pi R$ 
the top-squark wavefunction is required to change sign, while the 
top-quark wavefunction is invariant. These models illustrate the general 
properties of supersymmetry breaking that we require: at short 
distances, whether near the Yukawa brane or at a typical point 
in the bulk, all interactions are supersymmetric. The supersymmetric
cancellation in the Higgs mass calculation is prevented because the
top-multiplet KK modes feel supersymmetry breaking in their
wavefunctions on distance scales of order $R$. 
In sections \ref{section:localized} and \ref{section:Z2R-SS} 
we also compute the Higgs mass without performing a KK decomposition of 
the top multiplet, by studying the propagators of the top multiplet 
in position space in the bulk.  This calculation demonstrates the 
insensitivity of our results to ultraviolet physics --- one only needs 
a reliable effective theory at the energy scale $1/R$, since contributions 
from 4-momenta above this are exponentially damped. It also demonstrates 
the 5d supersymmetric cancellation more clearly than the KK mode calculation. 

In section \ref{section:Higgs} we discuss Higgs sectors on the 
Yukawa brane which successfully give masses to the Higgsinos and provide 
a restoring potential for the Higgs field. The phenomenology of our
theories is briefly studied in section \ref{section:pheno}, with 
emphasis on the different nature of the lightest superparticle (LSP) 
in the various models. Our conclusions are drawn in section 
\ref{section:conc}.

The model of section \ref{section:localized} also illustrates a new 
dynamical solution to the supersymmetric flavor problem. Even though 
squarks of the three generations have different couplings to the 
supersymmetry-breaking brane, the resulting squark wavefunctions are 
nearly identical, giving near degeneracy, as long as these couplings 
are all large.

\section{Radiative Correction to the Higgs-Boson Mass}
\label{section:rad-cor}

\subsection{Framework}

In this paper, we work in a framework of 5d supersymmetric models 
with the fifth dimension compactified on an $S^1/Z_2$ orbifold.  
The minimal supersymmetric multiplets in 5d are hypermultiplets and vector 
supermultiplets. The vector supermultiplet contains two Weyl fermions 
$\lambda_1$ and $\lambda_2$, a five-vector gauge field $A_M$, 
and a real scalar $\sigma$, all in the adjoint representation.  
The hypermultiplet contains two complex scalars $\phi$ and $\phi^c$ 
and two Weyl fermions $\psi$ and $\psi^c$.  Under the 4d $N=1$ 
supersymmetry, the vector supermultiplet fields form a vector superfield 
$V(\lambda_1,\; A_\mu)$ and an adjoint chiral superfield 
$\Sigma({1 \over \sqrt{2}}(\sigma+iA_5),\; \lambda_2)$, 
while the hypermultiplet fields form two chiral superfields, 
$\Phi(\phi,\; \psi)$ and $\Phi^c(\phi^c,\;\psi^c)$, with opposite 
quantum numbers. 

When we compactified the extra dimension on $S^1/Z_2$, we have two 
different types of fields; the bulk fields and the brane fields.
Let us label the coordinate $y$ of the extra dimension such that 
the orbifold fixed points are at $y=0$ and $y=\pi R$, where 
the $Z_2$ identifies points under the reflection $y \leftrightarrow -y$.
The bulk fields propagate in all five dimensions, and are classified 
according to whether they have an even or odd transformation 
under the $Z_2$ reflection.
Since the form of the Lagrangian requires that $\Phi$ and $\Phi^c$, 
and $V$ and $\Sigma$, have opposite transformation property, we take
\begin{eqnarray}
  && \Phi(x,-y)=\Phi(x,y), \qquad  \Phi^c(x,-y) =-\Phi^c(x,y),\\
  && V(x,-y)=V(x,y),       \qquad  \Sigma(x,-y)=-\Sigma(x,y). 
\end{eqnarray}
Note that the odd fields such as $\Phi^c$ and $\Sigma$ do not have 
zero modes after the KK decomposition.
Therefore, the orbifold compactification breaks the original 5d 
supersymmetry to 4d $N=1$ supersymmetry in the zero-mode sector.
On the other hand, the brane fields are localized on the orbifold 
fixed point, and can propagate only on the four dimensional hypersurface.
Thus, they form 4d $N=1$ supersymmetry multiplets and do not have
any KK towers after the KK decomposition.

Throughout the paper, we take all three generations of standard-model 
quarks and leptons, contained in hypermultiplets, and all standard-model 
gauge fields, contained in vector supermultiplets, to propagate in the 
extra dimension.  That is, chiral superfields $Q$, $U$, $D$, $L$,
and $E$ propagate in the bulk along with their conjugate superfields
$Q^c$, $U^c$, $D^c$, $L^c$, and $E^c$, and each vector superfield $V$ 
is accompanied by the corresponding chiral adjoint $\Sigma$.
The quark and lepton multiplets interact with the Higgs fields through 
the 4d $N=1$ supersymmetric Yukawa interactions located on the orbifold 
fixed point at $y=0$ (the Yukawa brane).
The two Higgs doublets $H_u$ and $H_d$ are required to give both 
up-type and down-type quark masses.  They can be either bulk or 
brane fields.  If the Higgs fields are the bulk fields, they are 
accompanied by the conjugate fields, $H_u^c$ and $H_d^c$.

Once supersymmetry is broken, the masses of the squark tower are shifted 
relative to those of the quark tower, and this effect is transmitted 
to the Higgs boson through radiative corrections.
Here we consider a class of models where the Higgs soft masses are 
zero at the tree level even after supersymmetry breaking.
Two explicit examples for such theories are given in section 
\ref{section:localized} and section \ref{section:Z2R-SS}.
Then, the Higgs scalars receive soft masses only radiatively through the 
loops of the bulk quark multiplets.  In the next subsection, we explicitly 
calculate the one-loop radiative correction to the up-type Higgs-boson 
mass coming from the loops of the KK towers of the top-quark 
hypermultiplets through the top-Yukawa coupling on the Yukawa brane.
We find that the result is finite in contrast to the usual 4d 
supersymmetric models where it is logarithmically divergent.
We also show that the correction is negative, so that it can indeed 
trigger EWSB.  A complete discussion of the Higgs sector will be postponed 
until section \ref{section:Higgs}.

\subsection{Calculation of the Higgs mass-squared}

In this subsection, we derive formulae for the radiatively generated 
Higgs-boson mass-squared, assuming that the Higgs boson is a brane field.
However, the final formulae written in terms of the 4d top-Yukawa 
coupling are also correct in the case where the Higgs boson 
is a bulk field, and are applicable in a class of extra 
dimensional theories discussed in sections \ref{section:localized} 
and \ref{section:Z2R-SS}.

We calculate the Higgs-boson mass by making a KK decomposition of the 
original 5d theory.
After the KK decomposition, the kinetic terms for the KK modes of the 
quark fields are written in terms of the canonically normalized fields as 
\begin{eqnarray}
  S_{\rm kin} &=&
  \int d^4 x \Biggl[ \Biggr\{
    \sum_{k=0}^{\infty} \Bigl(\partial^\mu \phi_{Q,k}^{\dagger} \partial_\mu 
    \phi_{Q,k} - m_{\phi_Q,k}^2 \phi_{Q,k}^{\dagger} \phi_{Q,k} \Bigr)
  + \sum_{k=1}^{\infty} \Bigl(\partial^\mu \phi^{c\,\dagger}_{Q,k} 
    \partial_\mu \phi^c_{Q,k} - m_{\phi_Q^c,k}^2 \phi^{c\,\dagger}_{Q,k} 
    \phi^c_{Q,k} \Bigr)
\nonumber\\
&&+ \sum_{k=1}^{\infty} \Bigl(\psi_{Q,k}^{\dagger}\, i\bar{\sigma}^\mu 
    \partial_\mu \psi_{Q,k} + \psi^{c\,\dagger}_{Q,k}\, i\bar{\sigma}^\mu 
    \partial_\mu \psi^c_{Q,k}
    - m_{\psi_Q,k} \psi^c_{Q,k} \psi_{Q,k} 
    - m_{\psi_Q,k} \psi_{Q,k}^{\dagger} \psi^{c\,\dagger}_{Q,k} \Bigr)
\nonumber\\
&&+ \psi_{Q,0}^{\dagger}\, i\bar{\sigma}^\mu \partial_\mu \psi_{Q,0} 
  \Biggr\} + \Bigl\{ Q \rightarrow U \Bigr\} \Biggr],
\end{eqnarray}
where $\phi_X$ and $\psi_X$ ($\phi^c_X$ and $\psi^c_X$) 
represent the scalar and fermion components of the chiral 
superfield $X$ ($X^c$), respectively.

The interactions between the Higgs fields and the KK modes of the quark 
fields are located on the $y=0$ fixed point:
\begin{equation}
  S_{\rm Yukawa} = \int\!\! d^4\! x\, dy \; \delta(y) 
  \left[-\int \!\!d^2\! \theta 
  \left({f_t \over M_*} Q_3 U_3 H_u 
      + {f_b \over M_*} Q_3 D_3 H_d + \cdots \right) 
  + {\rm h.c.} \right],
\label{eq:Yukawa-int}
\end{equation}
where the the chiral superfields $Q, U$ and $D$ are normalized in 5d 
so that they have mass dimension $3/2$, and $M_*$ is the cutoff 
of the theory.  Expanding the above brane interactions in 
component fields and eliminating the auxiliary $F$ fields, the
relevant interactions between canonically normalized 4d fields 
are found to be
\begin{eqnarray}
  S_{\rm int} &=&
  \int d^4 x \Biggl[
    \sum_{k=1}^{\infty} \sum_{l=0}^{\infty} 
    \Bigl(f_t \epsilon\, m_{\phi_Q^c,k} 
    \eta^{F_Q}_k \eta^{\phi_U}_l
    \phi^{c\,\dagger}_{Q,k} \phi_{U,l} \phi_H
  + f_t \epsilon\, m_{\phi_U^c,k} 
    \eta^{F_U}_k \eta^{\phi_Q}_l 
    \phi^{c\,\dagger}_{U,k} \phi_{Q,l} \phi_H
  + {\rm h.c.} \Bigr)
\nonumber\\
&&- \sum_{k=0}^{\infty} \sum_{l=0}^{\infty} \sum_{m=0}^{\infty} 
    \Bigl(f_t^2 \epsilon^2 
    \eta^{\phi_Q}_k \eta^{\phi_Q}_l (\eta^{F_U}_m)^2
    \phi^{\dagger}_{Q,k} \phi_{Q,l} \phi^{\dagger}_H \phi_H
  + f_t^2 \epsilon^2
    \eta^{\phi_U}_k \eta^{\phi_U}_l (\eta^{F_Q}_m)^2
    \phi^{\dagger}_{U,k} \phi_{U,l} \phi^{\dagger}_H \phi_H \Bigr)
\nonumber\\
&&- \sum_{k=0}^{\infty} \sum_{l=0}^{\infty} 
    \Bigl(f_t \epsilon\, \eta^{\psi_Q}_k \eta^{\psi_U}_l
    \psi_{Q,k} \psi_{U,l} \phi_H 
  + {\rm h.c.} \Bigr) \Biggr],
\end{eqnarray}
where $\epsilon$ is defined by $\epsilon \equiv 1/(\pi R M_*)$.
Here, $\eta^{\phi_X}_k$, $\eta^{\psi_X}_k$ and $\eta^{F_X}_k$ 
are the values of the wavefunctions at $y = 0$ for the $\phi_{X,k}$, 
$\psi_{X,k}$ and $F_{X,k}$ fields, respectively.

The Higgs-boson mass $m_{\phi_H}$ is generated at the one-loop level 
via loops of KK towers of the $Q$ and $U$ multiplets.
There are three types of diagrams, as shown in Fig.~\ref{Fig_RC-Higgs}, 
giving the three contributions
\begin{eqnarray}
-i\, m_{\phi_H}^2 &=& 
  N_c f^2 \epsilon^2 \sum_{k=1}^{\infty} \sum_{l=0}^{\infty}
    (\eta^{F_Q}_k)^2 (\eta^{\phi_U}_l)^2 
    \int \frac{d^4 p}{(2\pi)^4} 
    \frac{m_{\phi_Q^c,k}^2}{(p^2-m_{\phi_Q^c,k}^2)(p^2-m_{\phi_U,l}^2)}
    + (Q \leftrightarrow U)
\nonumber\\
&&-2 N_c f^2 \epsilon^2 \sum_{k=0}^{\infty} \sum_{l=0}^{\infty}
    (\eta^{\psi_Q}_k)^2 (\eta^{\psi_U}_l)^2 
    \int \frac{d^4 p}{(2\pi)^4} 
    \frac{p^2}{(p^2-m_{\psi_Q,k}^2)(p^2-m_{\psi_U,l}^2)}
\nonumber\\
&&+ N_c f^2 \epsilon^2 \sum_{k=0}^{\infty} \sum_{l=0}^{\infty}
    (\eta^{\phi_Q}_k)^2 (\eta^{F_U}_l)^2 
    \int \frac{d^4 p}{(2\pi)^4} 
    \frac{1}{(p^2-m_{\phi_Q,k}^2)}
    + (Q \leftrightarrow U).
\label{Higgs-gen}
\end{eqnarray}
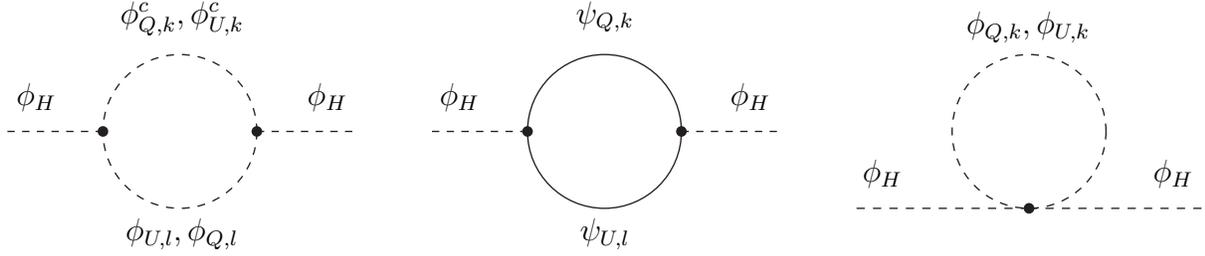
\begin{figure}
\begin{center} 
\begin{picture}(450,70)(-265,-35)
  \DashLine(-265,10)(-229,10){3} \Vertex(-229,10){2}
  \Text(-255,20)[b]{$\phi_H$}
  \DashCArc(-200,10)(29,0,360){3}
  \Text(-200,51)[b]{$\phi^c_{Q,k}, \phi^c_{U,k}$}
  \Text(-200,-23)[t]{$\phi_{U,l}, \phi_{Q,l}$}
  \DashLine(-171,10)(-135,10){3} \Vertex(-171,10){2}
  \Text(-145,20)[b]{$\phi_H$}
  \DashLine(-105,10)(-69,10){3} \Vertex(-69,10){2}
  \Text(-95,20)[b]{$\phi_H$}
  \CArc(-40,10)(29,0,360)
  \Text(-40,51)[b]{$\psi_{Q,k}$}
  \Text(-40,-23)[t]{$\psi_{U,l}$}
  \DashLine(-11,10)(25,10){3} \Vertex(-11,10){2}
  \Text(15,20)[b]{$\phi_H$}
  \DashLine(55,-19)(120,-19){3} \Text(65,-9)[b]{$\phi_H$}
  \DashCArc(120,10)(29,0,360){3} \Vertex(120,-19){2}
  \Text(120,47)[b]{$\phi_{Q,k}, \phi_{U,k}$}
  \DashLine(120,-19)(185,-19){3} \Text(175,-9)[b]{$\phi_H$}
\end{picture}
\caption{One-loop diagrams contributing to the Higgs-boson mass.}
\label{Fig_RC-Higgs}
\end{center}
\end{figure}
The exotic diagram, where the conjugate scalar fields 
$\phi^c_{Q,k}$ and $\phi^c_{U,k}$ circulate in the loop, 
is not present in the usual 4d supersymmetric models.
This diagram is actually needed to ensure the cancellation of 
$m_{\phi_H}^2$ in the supersymmetric limit, 
$m_{\phi_X,k} = m_{\phi_X^c,k} = m_{\psi_X,k}$ and 
$\eta^{\phi_X}_k = \eta^{\psi_X}_k = \eta^{F_X}_k$.

When supersymmetry is broken, the masses for the squark and quark 
towers shift relatively.
We here assume that the KK mass spectrum and wavefunctions do not 
depend on the species $Q$ and $U$, for simplicity, and drop the 
subscript $X$ from all quantities hereafter.
This assumption is indeed satisfied in a broad class of extra dimensional 
models, at least at the leading order in $1/(M_* R)$, including the two 
explicit models discussed later.
Supersymmetry breaking effects are represented by the 
deviations of the KK masses and wavefunctions from their supersymmetric 
relations, such as $m_{\phi,k} = m_{\psi,k}$ and 
$\eta^{\phi}_k = \eta^{\psi}_k$.
In general, the mass $m_k$ for the $k$-th KK excitation mode can be 
a rather complicated function of $k$; often it is even not equally 
spaced in $k$.  However, here we focus on some simple cases in which 
the masses for some of the KK towers are shifted by half a 
unit of $1/R$, and illustrate the basic idea that the Higgs-boson 
mass-squared receives a finite and negative contribution from the loops 
of the KK towers of the quark multiplets.
A more intuitive understanding of this remarkable property from the 5d 
viewpoint will be given in due course, in the context of explicit models.

We consider the following spectrum for the KK towers:
\begin{eqnarray}
  && m_{\psi,k}   = k \frac{1}{R} \qquad (k = 0,1,2,\cdots), \label{eq:mm1}\\
  && m_{\phi,k}   = (k + r^{\phi}) \frac{1}{R} \qquad (k = 0,1,2,\cdots), \\
  && m_{\phi^c,k} = (k + r^F) \frac{1}{R} \qquad 
    \left( \left\{ \begin{array}{ll}
      k = 1,2,3,\cdots \quad & {\rm for} \quad r^{F} = 0\\ 
      k = 0,1,2,\cdots \quad & {\rm for} \quad r^{F} = \frac{1}{2}\\ 
    \end{array} \right. \right),
\end{eqnarray}
where $r^{\phi}$ and $r^F$ take the value $0$ or $1/2$. 
The supersymmetric limit corresponds to $r^{\phi} = r^F = 0$.
Since the $r^{\phi} = 0$ case means massless squarks at the tree level
and is outside of the framework here,\footnote{In this case, the squarks 
obtain masses at the one-loop level through the standard-model 
gauge interactions, and the supersymmetry breaking is 
further transmitted to the Higgs boson at one more loop order 
through the top-Yukawa coupling.} we concentrate on the two 
cases $(r^{\phi}, r^F) = (1/2, 0)$ and $(1/2, 1/2)$ below.
These two cases corresponds to two explicit models discussed in 
sections \ref{section:localized} and \ref{section:Z2R-SS}.
The wavefunctions $g_k(y)$ for the KK modes are normalized such that 
$\int_0^{\pi R} [g_k(y)]^2 dy = \pi R$, giving 
\begin{eqnarray}
  && \eta^{\psi}_k = (\frac{1}{\sqrt{2}})^{\delta_{k,0}}, \\
  && \eta^{\phi}_k = 1, \\
  && \eta^{F}_k = \left\{ \begin{array}{ll}
    (\frac{1}{\sqrt{2}})^{\delta_{k,0}} 
    \qquad & {\rm for} \quad r^{F}=0,\\ 
    1 \qquad & {\rm for} \quad r^{F}=\frac{1}{2},\\ 
  \end{array} \right. \label{eq:ee3}
\end{eqnarray}
where $k = 0,1,2,\cdots$.
Plugging Eqs.~(\ref{eq:mm1} -- \ref{eq:ee3}) into a general 
expression Eq.~(\ref{Higgs-gen}), we obtain the radiative correction 
to the Higgs-boson mass.

Performing a Wick rotation to Euclidean momentum space $p_E$, 
and changing to the variable $x = p_E R$, gives
\begin{eqnarray}
-i\, m_{\phi_H}^2 &=& \frac{i N_c f_t^2 \epsilon^2}{R^2} 
    \int \frac{d^4 x}{(2\pi)^4} x^2 
\nonumber\\
&&  \times
    \sum_{k,l=0}^{\infty}
    \Biggl[ \frac{(\eta^{\psi}_k)^2 (\eta^{\psi}_l)^2}
      {(x^2+k^2)(x^2+l^2)} 
    - \frac{(\eta^{\phi}_k)^2 (\eta^{F}_l)^2}
      {(x^2+(k+r^{\phi})^2)(x^2+(l+r^{F})^2)} \Biggr].
\label{sum-mass}
\end{eqnarray}
In this expression, we first sum over the infinite tower of 
KK states and then perform the momentum integral to obtain a 
sensible result.  
Given  that higher dimensional theories are non-renormalizable and must 
be cut off at some scale,
one might worry that summing up infinite KK states 
is not the correct procedure.
However, the point is that any cutoff must preserve the correct 
symmetries of the theory; 5d Lorentz symmetry and supersymmetry.
This is precisely what is done by summing up infinite towers of the 
KK states, and is difficult to attain in any other way.
In that sense, we can view this summation procedure as a kind of 
regularization, ``KK regularization''.
Indeed, after the summation, the resulting momentum integral turns out 
to be strongly dominated by the $1/R$ scale and the contribution 
from near the 
cutoff scale is extremely small.  This is also consistent with the 
5d picture that the Higgs boson requires some non-local information 
over the $1/R$ scale to feel supersymmetry breaking, which we will 
explicitly see in later sections.

Now, let us evaluate the Higgs-boson mass using Eq.~(\ref{sum-mass}).
It is easy to check that the expression vanishes in the supersymmetric 
limit $r^{\phi} = r^F = 0$.  In the case of $(r^{\phi}, r^F) = (1/2, 0)$, 
the Higgs-boson mass is given by
\begin{eqnarray}
  m_{\phi_H}^2 &=&
  - \frac{N_c f_t^2 \epsilon^2}{16R^2}
    \int_0^{\infty}dx \frac{x^3}{\sinh^2[\pi x]}  
\\
  &=& - \frac{3\,\zeta(3)}{32 \pi^4} \frac{N_c f_t^2 \epsilon^2}{R^2},
\label{Higgs-mass-1}
\end{eqnarray}
where $\zeta(x)$ is the Riemann's zeta function.
We find that the radiative correction $m_{\phi_H}^2$ is negative 
and EWSB is indeed triggered by the loops of the KK towers.
Furthermore, the result is finite and ultraviolet insensitive; the 
momentum integral is exponentially cut off at $p_E \sim (\pi R)^{-1}$ 
as was promised earlier.
This extreme softness seems to come from a miraculous cancellation 
between fermionic and bosonic KK modes from the 4d point of view. There 
is a beautiful understanding of this result from the 5d 
viewpoint, which will be discussed in the context of an explicit model 
in section \ref{section:localized}, and more generally in the conclusion.

In the $(r^{\phi}, r^F) = (1/2, 1/2)$ case, the Higgs mass-squared is
\begin{eqnarray}
  m_{\phi_H}^2 &=&
  - \frac{N_c f_t^2 \epsilon^2}{16R^2} \int_0^{\infty}dx\, x^3 
  \left\{ \coth^2 [\pi x] - \tanh^{2}[\pi x] \right\}
\\
  &=& - \frac{21\,\zeta(3)}{128 \pi^4} \frac{N_c f_t^2 \epsilon^2}{R^2}.
\label{Higgs-mass-2}
\end{eqnarray}
As in the case of $(r^{\phi}, r^F) = (1/2, 0)$, the radiative correction 
is negative and the momentum integral is exponentially cut off at 
$p_E \sim (\pi R)^{-1}$.  This case occurs in the model given in 
section \ref{section:Z2R-SS}, where we also discuss a 5d interpretation 
of the result.

Finally, we can rewrite the expressions in Eqs.~(\ref{Higgs-mass-1}, 
\ref{Higgs-mass-2}) in terms of the 4d top-Yukawa coupling 
$y_t = f_t \epsilon / 2$.  They are given by
\begin{eqnarray}
  m_{\phi_H}^2 &=&
  - \frac{3\,\zeta(3)}{8 \pi^4} \frac{N_c\, y_t^2}{R^2} 
\label{mH-1-1}\\
  &\simeq& -(1.39 \times 10^{-2}) \frac{y_t^2}{R^2},
\label{mH-1-2}
\end{eqnarray}
for $(r^{\phi}, r^F) = (1/2, 0)$ and 
\begin{eqnarray}
  m_{\phi_H}^2 &=&
  - \frac{21\,\zeta(3)}{32 \pi^4} \frac{N_c\, y_t^2}{R^2} 
\label{mH-2-1}\\
  &\simeq& -(2.43 \times 10^{-2}) \frac{y_t^2}{R^2},
\label{mH-2-2}
\end{eqnarray}
for $(r^{\phi}, r^F) = (1/2, 1/2)$.   Here, we have used $N_c=3$ and 
$\zeta(3) \simeq 1.202$.   Note that although the above results in 
Eqs.~(\ref{mH-1-1} -- \ref{mH-2-2}) are derived by assuming that 
the Higgs boson is a brane field, they are also valid in the case of 
the bulk Higgs field.  This can be easily verified by carefully tracing 
the volume-suppression and wavefunction-normalization factors coming 
from the zero-mode Higgs boson.

The resulting values of $m_{\phi_H}^2$ are one-loop suppressed 
compared with $(1/R)^2$.  This means that the superparticle masses 
can be naturally larger than the weak scale in contrast to the 
usual 4d supersymmetric models.
If we consider $m_{\phi_H} \simeq 300~{\rm GeV}$, for instance, 
the compactification scale $R^{-1}$ can be as high as 
$R^{-1} \simeq (2 \sim 3)~{\rm TeV}$, corresponding to the squark mass 
$(2 R)^{-1} \simeq (1.0 \sim 1.5)~{\rm TeV}$.   This hierarchy between 
the squark and the Higgs-boson masses is a consequence of the fact 
that the Higgs soft mass is zero at the tree level.

\section{Model with Localized Supersymmetry Breaking}
\label{section:localized}

In this and the next sections, we discuss two explicit models which 
realize the form of supersymmetry breaking discussed in the previous section.
The two models have quite different mechanisms of realizing the desired 
properties: vanishing Higgs soft masses at the tree level and 
finiteness of the radiative correction to the Higgs-boson mass.
We also give a useful physical picture to understand those properties 
in each model.

\subsection{Setup}
\label{subsection:localsetup}

The first model we consider has the following structure.
Let us consider two branes located at two orbifold fixed points, 
$y=0$ and $\pi R$.  The two Higgs-doublet chiral superfields $H_u$ and 
$H_d$ are localized on the $y=0$ brane and the supersymmetry breaking 
occurs on the other brane at $y=\pi R$.   A distinctive feature of the 
present model is that supersymmetry is strongly broken at the $y=\pi R$ 
fixed point by $\langle Z \rangle \sim M_*^2 \theta^2$, where $Z$ is 
a chiral superfield localized at the fixed point and $M_*$ is the 
cutoff of the theory.  This is easily realized by considering the brane 
superpotential
\begin{equation}
  S_Z = \int\!\! d^4\! x\, dy \; \delta(y-\pi R) 
  \left[ \int \!\!d^2\! \theta M_*^2 Z + {\rm h.c.} \right].
\end{equation}
With this strong supersymmetry breaking, the bulk fields such as quark, 
lepton and gauge multiplets feel the supersymmetry breaking effect 
through the interactions with the $Z$ field.

The bulk fields and the brane fields confined on the orbifold fixed 
points can have localized interactions that respect the 4d $N=1$ 
supersymmetry.  Thus, the bulk fields feel the supersymmetry breaking 
through the following interactions:
\begin{eqnarray}
  S_{\pi R} &=& \int\!\!d^4\!x\, dy \; \delta(y-\pi R) \left[ 
    -\int \!\!d^4\! \theta \left( {c_Q \over M_*^3}Q^\dagger Q Z^\dagger Z
    +{c_U \over M_*^3} U^\dagger U Z^\dagger Z + \cdots \right)\right. 
\nonumber \\
  && \qquad\qquad\qquad\qquad + \left.\left(\int \!\!d^2\!\theta 
    {1 \over 16 g_5^2} {c_W \over M_*^2} Z\, {\rm Tr}W^\alpha W_\alpha
    + {\rm h.c.} \right)\right].
\label{eq:ssbint}
\end{eqnarray}
Here the $c$'s are dimensionless, and $g_5$, the 5d gauge coupling,
has mass dimension $-1/2$.  We assume that the coupling constants $c$ 
are all positive.  The wavefunctions of the fields that feel the 
supersymmetry breaking (the squarks, sleptons, and gauginos) will be 
repelled from the $y=\pi R$ fixed point, making their zero modes massive.  
This effect will be examined in detail in subsection 
\ref{subsection:localspectrum}.

The wavefunctions for fields odd under the $Z_2$ orbifold
are forced to vanish at the fixed points, so that interactions like
\begin{equation}
  \int\!\!d^4\!x\, dy \; \delta(y-\pi R) \int \!\!d^4
  \! \theta   {1\over M_*^3}{Q^c}^\dagger Q^c Z^\dagger Z,
\end{equation}
do not arise.  Derivative interactions such as
\begin{equation}
  \int\!\!d^4\!x\, dy \; \delta(y-\pi R) \int \!\!d^4
  \! \theta   {1\over M_*^5} \partial_y {Q^c}^\dagger 
  \partial_y Q^c Z^\dagger Z,
\end{equation}
can exist, but will be suppressed by factors of $1/(R M_*)$ for the lowest 
KK modes.  From an effective field theory standpoint, it is completely 
consistent to couple only $\Phi$ and $V$ superfields to the orbifold fixed
points and not $\Phi^c$ and $\Sigma$.  It is nevertheless interesting 
that even if one includes couplings of the odd fields, they are naturally 
suppressed. In what follows we will for simplicity ignore the possible 
effects of supersymmetry breaking on the wavefunctions of $\Phi^c$ 
and $\Sigma$ fields.  These effects, if present, would be smaller than 
those for the even field wavefunctions at lower KK modes, so the 
qualitative results we obtain should apply regardless.

Another important feature of the present model is that the Higgs chiral 
multiplets do not acquire a tree-level supersymmetry-breaking mass, 
since they are localized on the brane at $y = 0$ and do not have any 
direct interactions with the $Z$ field.\footnote{
There may be some direct couplings between the Higgs and $Z$ fields 
generated by exchange of heavy fields of masses of the order of the 
cutoff scale.  However, they are exponentially suppressed as 
$\exp(-\pi M_* R)$, so that contributions from these operators are 
sufficiently small compared with the one-loop contribution 
calculated in section \ref{section:rad-cor}, if $M_* R \gsim 2$.}
Therefore, they feel the supersymmetry breaking only radiatively through 
their couplings to the even bulk fields given in Eq.~(\ref{eq:Yukawa-int}).
Our model depends crucially on the Higgs soft mass being zero at
the tree level, so that $m_{H_u}^2$ is driven negative by the large 
top-Yukawa coupling.  In subsection \ref{subsection:localspectrum} we will 
obtain the tree-level spectrum for the KK modes of the bulk fields, which 
will enable us to compute this radiative effect using the formulae of 
section \ref{section:rad-cor}.  A complete discussion of the Higgs sector 
of our theory will be given in section \ref{section:Higgs}.

\subsection{Particle spectrum and radiatively induced $m_{H_u}^2$}
\label{subsection:localspectrum}

Neither $\psi$ nor $\psi^c$ couples to the supersymmetry breaking, so 
they have the expansions
\begin{equation}
  \psi(x,y)=\sum_{n=0}^\infty 
    \frac{1}{\sqrt{2}^{\delta_{n,0}}} {\cos[ny/R] \over
    \sqrt{\pi R}}\psi_n(x), \qquad 
  \psi^c(x,y)=\sum_{n=1}^\infty {\sin[ny/R] \over
    \sqrt{\pi R}}\psi^c_n(x).
\end{equation}
The expansion of $A_\mu$ is completely analogous to that of $\psi$, 
and the expansions of the $Z_2$-odd scalars $\phi^c$ and $\sigma$ are
analogous to that of $\psi^c$.
Defining $M_c=1/R$, the masses of the KK modes are
\begin{equation}
  m_{\psi,n}=m_{A^\mu,n}=nM_c \qquad (n=0,1,2,\cdots),
\label{eq:evenspectrum}
\end{equation}
and
\begin{equation}
  m_{\psi^c,n}=m_{\phi^c,n}=m_{\sigma,n}=nM_c \qquad (n=1,2,\cdots).
\label{eq:oddspectrum}
\end{equation}
For $n>0$, $\psi_n$ and $\psi_n^c$ marry to form a Dirac particle of
mass $n M_c$.  Note also that the non-zero modes of $A_\mu$ acquire mass by
eating the corresponding modes of the odd scalar $A_5$.  

The situation is more complicated for $\phi$ scalars because of
their coupling to $F_Z$.  Their classical equation of motion is
\begin{equation}
  \partial_{\mu}^2 \phi - \partial_y^2 \phi
  + \delta(y-\pi R) {c_X {F_Z}^2 \over M_*^3}\phi=0,
\end{equation}
where $X = Q, U, \cdots$.  Remembering that $\phi$ is even under the $Z_2$, 
the solution (for $0<y<\pi R$) is
\begin{equation}
  \phi(x,y) = \sum_{n=0}^\infty C_n \cos[m_{\phi,n} y] \phi_n(x), 
\end{equation}
where the $C_n$ are constants chosen to canonically normalize $\phi_n$ 
kinetic terms, and the 4d masses $m_{\phi,n}$ are the solutions to 
the equation
\begin{equation}
  \tan[m_{\phi,n} \pi R] = {c_X \over 2}{{F_Z}^2 \over M_*^4}
  {M_* \over m_{\phi,n}}.
\end{equation}
Taking $c_X \sim 1$ and $\sqrt{F_Z} \sim M_*$, one obtains
\begin{equation}
  m_{\phi,n} \simeq \left(n+\frac{1}{2}\right)M_c \qquad (n=0,1,2,\cdots),
\label{eq:phispectrum}
\end{equation}
at the leading order in $M_c/M_*$.  We see that at this order the 
supersymmetry breaking acts as an impenetrable wall that drives the 
wavefunction of $\phi_n$ to zero at $y=\pi R$, making the masses 
insensitive to the precise values of $c_X$ and $F_Z$.  
At the next order in $M_c/M_*$, one finds
\begin{equation}
  m_{\phi,n} \simeq \left(n+\frac{1}{2}\right)M_c 
  \left(1-{2 \over c_X \pi}{M_*^4\over {F_Z}^2}
  {M_c \over M_*}\right),
\label{eq:scalar-corr}
\end{equation}
giving a small but finite $\phi$ wavefunction at the 
supersymmetry-breaking brane.  

The Weyl fermions $\lambda_1$ and $\lambda_2$ are coupled through their
kinetic term, while only the even field $\lambda_1$ feels the supersymmetry 
breaking directly.  The classical equations of motion are
\begin{equation}
  -i{\overline \sigma}^\mu \partial_\mu \lambda_2
  +\partial_y{\overline \lambda}_1=0,
\end{equation}
\begin{equation}
  -i{\overline \sigma}^\mu \partial_\mu \lambda_1
  -\partial_y{\overline \lambda}_2
  -\delta(y-\pi R){c_W F_Z \over 2 M_*^2}{\overline \lambda_1}=0.
\end{equation}
Looking for solutions of the form 
\begin{equation}
  \lambda_j (x,y) = \eta_j(x) g_j(y) \qquad (j = 1,2),
\end{equation}
we find the boundary condition at the $y=\pi R$ fixed point
\begin{equation}
  \eta_2(x)={c_W F_Z \over 4 M_*^2}{g_1(\pi R) \over g_2(\pi R)}\, \eta_1(x).
\end{equation}
Setting $i{\overline \sigma}^\mu \partial_\mu \eta_1
= m_\lambda {\overline \eta_1}$ then leads to the solution 
\begin{equation}
  g_1\propto \cos[m_\lambda y], \qquad\qquad g_2\propto \sin[m_\lambda y],
\end{equation}
with the KK-mode masses given by
\begin{equation}
  \tan[m_\lambda \pi R] = {c_W F_Z \over 4 M_*^2}.
\label{eq:fermionmass}
\end{equation}
Note that Eq.~(\ref{eq:fermionmass}) has solutions for both
positive and negative $m_\lambda$ (the absolute value gives the physical
mass).  For instance, in the case of extremely weak supersymmetry breaking,
with $\epsilon \equiv c_W F_Z/(4 \pi M_*^2) \ll 1$, the masses are 
given by $\epsilon M_c$, $(1 \pm \epsilon)M_c$, $(2\pm \epsilon)M_c$, 
and so on.   For the opposite case of very strong supersymmetry breaking, 
where $\delta \equiv 4 M_*^2/(\pi c_W F_Z) \ll 1$, the masses are 
$(1/2 \pm \delta)M_c$, $(3/2 \pm \delta)M_c$, etc.

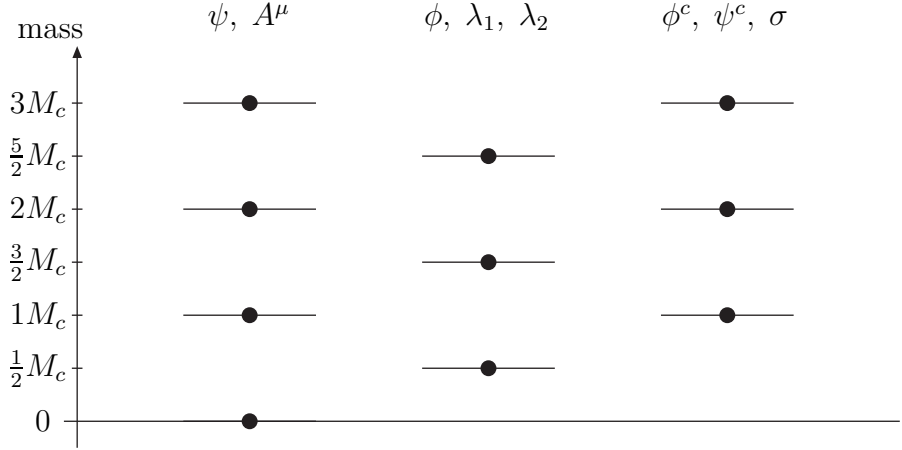
\begin{figure}
\begin{center} 
\begin{picture}(350,190)(-10,-25)
  \Line(5,0)(320,0)
  \LongArrow(10,-10)(10,140)
  \Text(0,145)[b]{mass}
  \Text(0,0)[r]{$0$}
  \Line(8,20)(12,20)    \Text(6,20)[r]{${1\over 2}M_c$}
  \Line(8,40)(12,40)    \Text(6,40)[r]{$1 M_c$}
  \Line(8,60)(12,60)    \Text(6,60)[r]{${3 \over 2}M_c$}
  \Line(8,80)(12,80)    \Text(6,80)[r]{$2 M_c$}
  \Line(8,100)(12,100)  \Text(6,100)[r]{${5 \over 2} M_c$}
  \Line(8,120)(12,120)  \Text(6,120)[r]{$3 M_c$}
  \Text(75,150)[b]{$\psi,\; A^{\mu}$}
  \Line(50,0)(100,0)      \Vertex(75,0){3}
  \Line(50,40)(100,40)    \Vertex(75,40){3}
  \Line(50,80)(100,80)    \Vertex(75,80){3}
  \Line(50,120)(100,120)  \Vertex(75,120){3}
  \Text(165,150)[b]{$\phi,\; \lambda_1,\;\lambda_2$}
  \Line(140,20)(190,20)    \Vertex(165,20){3}
  \Line(140,60)(190,60)    \Vertex(165,60){3}
  \Line(140,100)(190,100)  \Vertex(165,100){3}
  \Text(255,150)[b]{$\phi^{c},\; \psi^{c},\; \sigma$}
  \Line(230,40)(280,40)    \Vertex(255,40){3}
  \Line(230,80)(280,80)    \Vertex(255,80){3}
  \Line(230,120)(280,120)  \Vertex(255,120){3}
\end{picture}
\caption{Mass spectrum for the lowest KK modes of the bulk fields in
  our model, in the limit of very strong supersymmetry breaking, 
  $\sqrt{F_Z} \gg M_* \gg M_c$.  The fermion modes $\psi_n$ and 
  $\psi^c_n$ form a Dirac state for $n>0$, while the modes of 
  ${\lambda_1}$ and ${\lambda_2}$ combine
  to form nearly degenerate pairs of Majorana states.}
\label{fig:spectrum-1}
\end{center}
\end{figure}
The spectrum of states is summarized in Fig.~\ref{fig:spectrum-1} in the
limit $\sqrt{F_Z} \gg M_* \gg M_c$.  Lowering $\sqrt{F_Z}$
down to $M_*$ has only a small effect of order $M_c/M_*$ on the $\phi_n$ 
masses.  In contrast, changes of order unity arise in the gaugino mass 
spectrum, unless $c_W \gg 1$.  In particular, the near-degeneracy 
between pairs of their KK modes that exists in the strong 
supersymmetry-breaking limit is spoiled. 

Knowing the KK expansions of $\psi$, $\phi$, and $\phi^c$ allows us 
to compute the one-loop induced mass-squared for the Higgs boson 
by applying the result of section \ref{section:rad-cor}.
Using Eqs.~(\ref{eq:evenspectrum}, \ref{eq:oddspectrum}, 
\ref{eq:phispectrum}), and taking $\eta_k^\phi=1$ and 
$\eta_k^F=\eta_k^\psi=(1/\sqrt{2})^{\delta_{k,0}}$ accordingly,
we obtain the up-type Higgs-boson mass 
\begin{equation}
  m_{\phi_{H_u}}^2 = - {3\, \zeta(3) \over 8 \pi^4} N_c\, y_t^2 M_c^2,
\label{eq:HG-1}
\end{equation}
from Eq.~(\ref{mH-1-1}), since the KK mass spectrum in 
Fig.~\ref{fig:spectrum-1} corresponds to the $(r^{\phi}, r^F) = (1/2, 0)$ 
case in section \ref{section:rad-cor}.
We have also checked that the $O(M_c/M_*)$ correction in the scalar mass 
in Eq.~(\ref{eq:scalar-corr}) gives only $O(M_c/M_*)$ correction to 
the Higgs-boson mass and can be safely neglected.

\subsection{5d interpretation}

It is remarkable that the Higgs mass is not merely ultraviolet insensitive, 
but in fact ultraviolet finite. One might, however, be skeptical of this 
result. For instance, we have summed over {\em all} modes 
in the KK expansion, rather than merely those below the 
cutoff. If we were to introduce an explicit cutoff, a strong sensitivity 
to this cutoff would be present.  Ordinarily a top-stop pair contributes 
to the Higgs mass an amount $\sim m^{2}_{\rm SUSY}/(16 \pi^{2})$. Now 
that we have a tower, we would expect a multiplicity $N_{\rm KK}^2 \simeq 
(M_{*} R)^2$ as well, which is what we find with an explicit 
cutoff. This result is incorrect, because when we sum the entire 
tower, we are in fact Fourier transforming to mixed position-momentum 
space \cite{Arkani-Hamed:2000za}. In this formulation it is clear that 
the result {\em must} be finite, and the total summation of all modes is 
the proper thing to do. 

Since the Yukawa brane is located at $y=0$ and the supersymmetry 
breaking is located at $y=\pi R$, for values of momenta $k_{4}>(\pi 
R)^{-1}$ the Higgs boson cannot simultaneously ``see'' both the Yukawa 
couplings and the supersymmetry breaking. Since the contribution to 
the Higgs mass relies on both of these, it must vanish exponentially 
at high momenta, just as we have seen in the previous calculation.
This very intuitive result is masked by the KK formalism, 
despite its calculational utility. Here we will calculate in the 
mixed position-momentum space from the outset and these results will 
appear quite naturally.

To begin with, we must calculate the scalar propagator with the 
boundary mass term. We shall consider first the case of an infinite 
dimension before dealing with the compact case. We will work from 
the outset in Euclidean space. The equation for the 
Green's function $G_{\phi}$ is
\begin{equation}
  (-\partial_{4}^{2}-\partial_{y}^{2}) G_{\phi}(x_{4},y) + 
  m \delta(y-\pi R) G_{\phi}(x_{4},y) = \delta^{4}(x_{4}) \delta(y),
\label{eq:kgequation}
\end{equation}
where $m \equiv c_X F_Z^2/M_*^3$.
Transforming to the mixed position-momentum space gives
\begin{equation}
  (k_{4}^{2}-\partial_{y}^{2}) \tilde G_{\phi}(k_{4},y) = 
  -m \delta(y-\pi R) \tilde G_{\phi}(k_{4},\pi R) +  \delta(y).
\end{equation}
This is simply the equation for the one dimensional propagator of a 
field with mass $M=k_{4}$ and sources at $y=0$ and $y=\pi R$ with 
strengths one and $-m \tilde G_{\phi}(k_{4},\pi R)$, respectively. 
Knowing this, we can explicitly write the solution:
\begin{equation}
  \tilde G_{\phi}(k_{4},y) = \frac{e^{-k_{4} |y|}}{2 k_{4}} - m \tilde 
  G_{\phi}(k_{4},\pi R) \frac{e^{-k_{4} |y-\pi R|}}{2 k_{4}}.
\end{equation}
We can solve for $\tilde G_{\phi}(k_{4},\pi R)$ and substitute back 
to get the complete result
\begin{equation}
  \tilde G_{\phi}(k_{4},y) = \frac{e^{-k_{4} |y|}}{2 k_{4}} 
  - m \frac{e^{-k_{4} \pi R - k_{4}|y-\pi R|}}{(2 k_{4})(2k_{4}+m)}.
\end{equation}
As before we can take the $m \rightarrow \infty$ limit and get
\begin{equation}
  \tilde G_{\phi}(k_{4},y) = \frac{e^{-k_{4} |y|}}{2 k_{4}}
  -\frac{e^{-k_{4} 2 \pi R + k_{4} y}}{2 k_{4}},
\end{equation}
for $y<\pi R$ and
\begin{equation}
  \tilde G_{\phi}(k_{4},y) = 0,
\end{equation}
for $y>\pi R$. Previously, we lacked an intuitive understanding of 
the $m\rightarrow \infty$ limit, where the result was insensitive to 
the value of $m$ so long as it was sufficiently high. Here we 
see that the supersymmetry-breaking brane reflects the $\phi$ field 
back and $m$ is a measure of the opacity. In the $m\rightarrow \infty$ 
limit, the wall becomes completely opaque, and the reflected signal is 
maximal.

\begin{figure}
  \centerline{
  \psfig{file=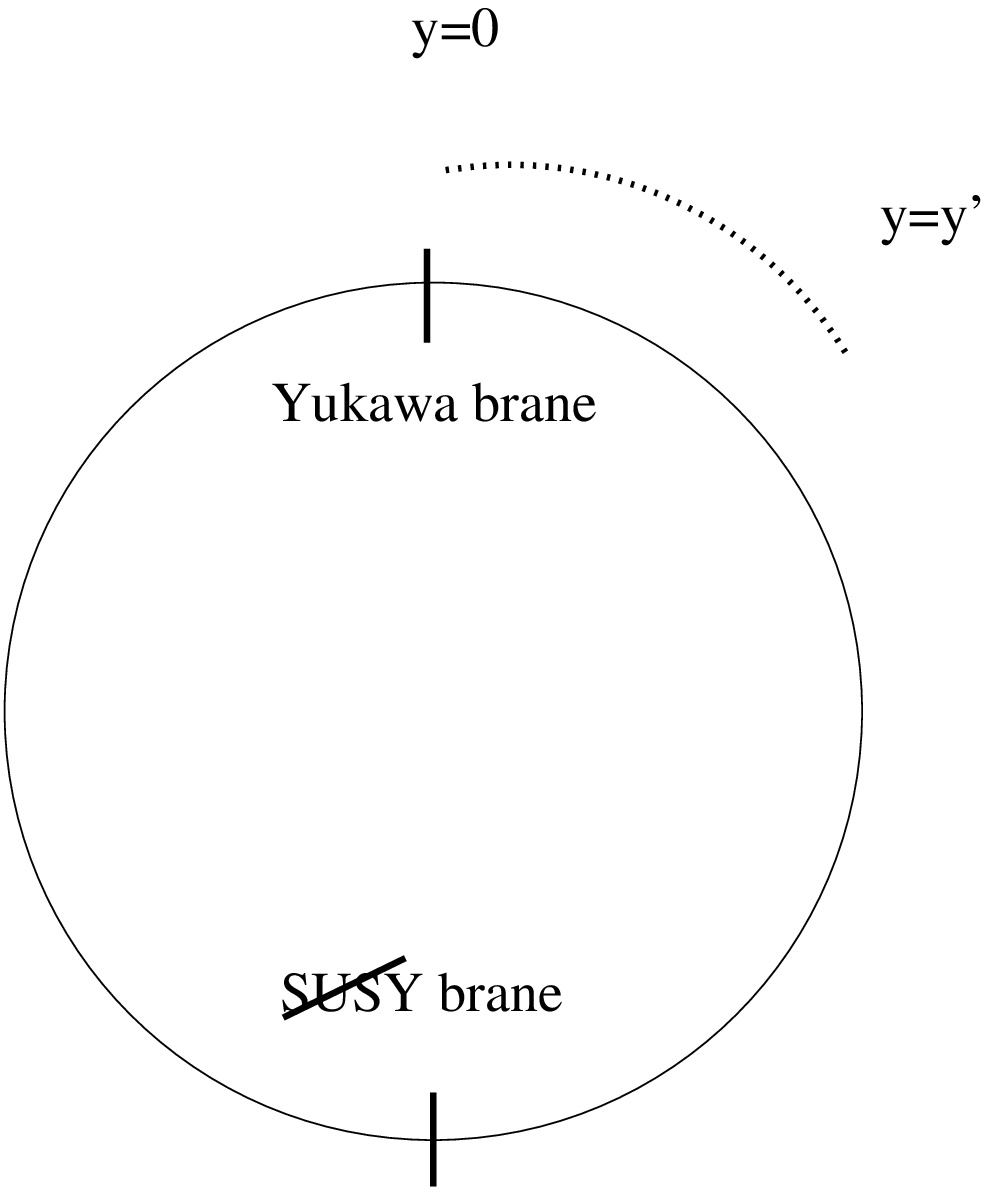,width=0.34\textwidth}\hskip0.1in
  \psfig{file=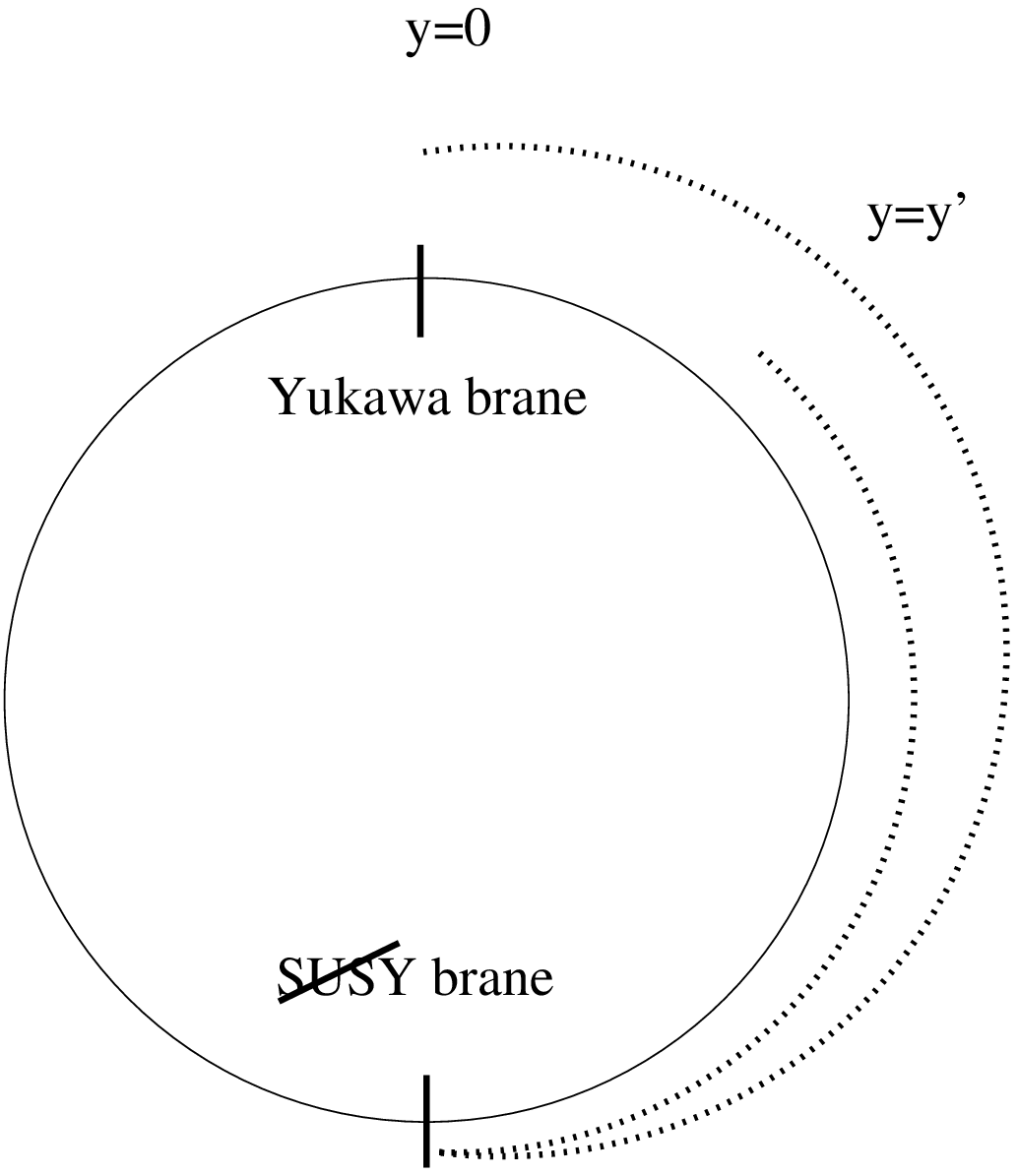,width=0.35\textwidth}\hskip0.1in
  \psfig{file=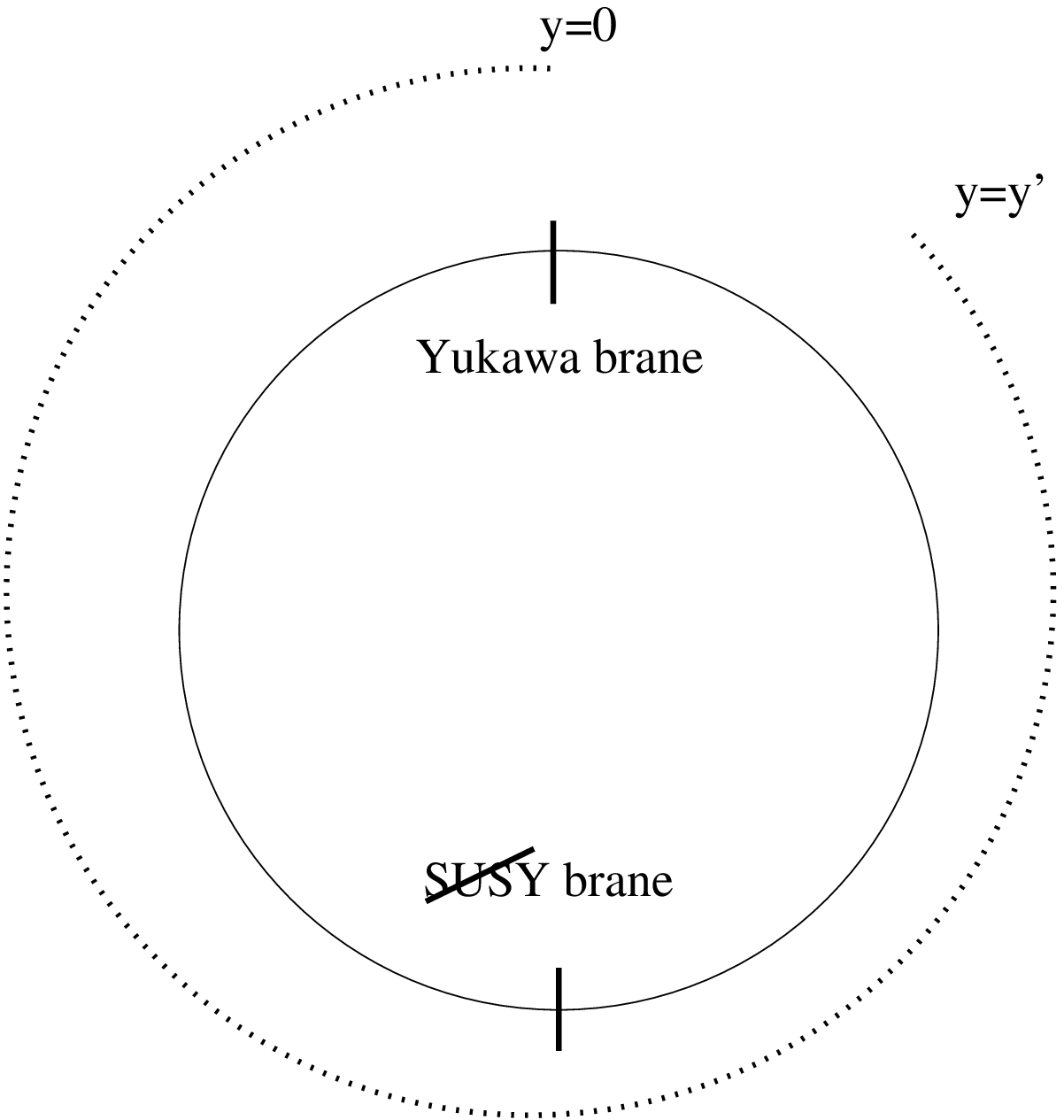,width=0.38\textwidth}}
\caption{Different contributions to the propagator from $y=0$ to 
  $y=y'$. In addition to the lowest order piece, there are both 
  reflected contributions and winding contributions. In the 
  $m\rightarrow \infty$ limit, the supersymmetry-breaking brane 
  becomes opaque and the winding contributions vanish, although 
  an infinite number of reflections contribute.}
\label{fig:propcon}
\end{figure}

With this understanding, we can address the compact dimension case. 
The proper generalization of Eq.~(\ref{eq:kgequation}) is 
\begin{equation}
  (-\partial_{4}^{2}-\partial_{y}^{2}) G_{\phi}(x_{4},y) 
  + m \sum_{n} \delta(y-(2n+1)\pi R) 
  G_{\phi}(x_{4},y) = \sum_{n}\delta^{4}(x_{4}) \delta(y-2 n \pi R),
\label{eq:kgequationcompact}
\end{equation}
where the summations now represent all possible windings, 
$n = -\infty, \cdots, \infty$.  Then, using image charge techniques, 
one can calculate the propagator in complete analog with the infinite 
dimension case, finding
\begin{equation}
  \tilde G_{\phi}(k_{4},y) = \frac{1}{2 k_{4}} 
  \frac{1}{\sinh[k_4 \pi R]} \left\{ \cosh[k_4 (\pi R - y)] - 
  \frac{m \cosh[k_4 y]}{2 k_4 \sinh[k_4 \pi R] + m \cosh[k_4 \pi R]}
  \right\},
\label{eq:genprop}
\end{equation}
for $(y \in [0, \pi R])$.
Again, we can take the opaque limit $m \rightarrow \infty$ and then 
the propagator becomes
\begin{equation}
  \tilde G_{\phi}(k_{4},y) = \frac{1}{2 k_{4}} 
  \frac{1}{\sinh[k_4 \pi R]} \left\{ \cosh[k_4 (\pi R - y)] - 
  \frac{\cosh[k_4 y]}{\cosh[k_4 \pi R]} \right\}.
\label{eq:opaprop}
\end{equation}
It is instructive to expand Eq.~(\ref{eq:opaprop}) in exponentials as
\begin{equation}
  \tilde G_{\phi}(k_{4},y) = \frac{1}{2 k_{4}} e^{- k_4 y}
  - \frac{1}{2 k_{4}} 
  \left( e^{- k_4 (2\pi R - y)} + e^{- k_4 (2\pi R + y)} \right)
  \left( 1 - e^{- k_4 2\pi R} + e^{- k_4 4\pi R} - \cdots \right),
\end{equation}
where the first term represents direct propagation and 
the second term represents contributions from reflections.
In this expression, we can explicitly see that reflection 
from the supersymmetry-breaking brane just 
gives a minus sign: the reflection does not reduce the strength of 
the signal, implying that the supersymmetry-breaking brane is an ideal 
mirror in the limit $m \rightarrow \infty$.
Note that we need not have resorted to solving 
Eq.~(\ref{eq:kgequationcompact}) to get the propagator in 
Eq.~(\ref{eq:opaprop}).   Since the wall is opaque, we could have 
found this result by summing all possible reflections 
from walls at $\pm \pi R$, picking up a minus sign at each reflection.  
The contributions to the propagator 
are represented in Fig.~\ref{fig:propcon}.

We also need the appropriate propagators for the $F$ components and 
fermions. Neither of the fields feel supersymmetry breaking directly, 
however, so it is relatively simple to calculate these.

For the $F$-component, we need only investigate the kinetic piece of 
the Lagrangian
\begin{equation}
  {\cal L} \supset \pmatrix{\phi^{c*} & F} \pmatrix{k_{4}^{2}& i 
  k_{y} \cr i k_{y} & 1} \pmatrix{\phi^{c} \cr F^{*}}.
\end{equation}
We can invert this to yield
\begin{equation}
  \pmatrix{k_{4}^{2}& i k_{y} \cr i k_{y} & 1}^{-1} = 
  \frac{1}{k_{4}^{2}+k_{y}^{2}}\pmatrix{1 & -i k_{y}\cr 
  -i k_{y} & k_{4}^{2}}.
\end{equation}
Our $F$-$F$ propagator is just $k_{4}^{2}/(k_{4}^{2}+k_{y}^{2})$, 
which we trivially Fourier transform to mixed position-momentum space 
to give
\begin{equation}
  \tilde G_{F}(k_{4},y) = \frac{k_{4}}{2}e^{- k_{4}|y|},
\end{equation}
in infinite space. Summing over winding modes, this becomes 
$(y \in [0,\pi R])$
\begin{equation}
  \tilde G_{F}(k_{4},y) = \frac{k_{4}}{2} 
  \frac{\cosh[k_{4}(\pi R- y)]}{\sinh[k_{4}\pi R]},
\label{eq:prop-F}
\end{equation}
for the compact space.

The fermion propagator is more straightforward. In momentum space 
the $\psi$-$\psi$ (fermion) propagator is just\footnote{There is also 
a piece proportional to $\gamma_{5} k_{y}$, but this connects the 
fermion to the conjugate fermion, which has no Yukawa couplings.}
\begin{equation}
  \tilde {\tilde G}_{\psi}(k_{4},k_{5})=
  \frac{\not k_{4}}{k_{4}^{2}+k_{y}^{2}},
\end{equation}
which we Fourier transform to yield
\begin{equation}
  \tilde G_{\psi}(k_{4},y) = \frac{\not k_{4}}{2 k_{4}}e^{- k_{4}|y|}.
\end{equation}
Again, including all windings, this becomes $(y \in [0, \pi R])$
\begin{equation}
  \tilde G_{\psi}(k_{4},y) = \frac{\not k_{4}}{2 k_{4}} 
  \frac{\cosh[k_{4}(\pi R- y)]}{\sinh[k_{4}\pi R]}.
\label{eq:prop-fermion}
\end{equation}
Eqs.~(\ref{eq:genprop}, \ref{eq:prop-F}, \ref{eq:prop-fermion}) 
provide all the propagators needed to calculate the Higgs-boson mass.

As for the interaction, we write our superpotential term in terms 
of the physical quantities
\begin{equation}
  W \supset \int dy\, \delta(y)\, 2\pi R\, y_{t}\, Q U H_{u}.
\end{equation}
In the Lagrangian this contributes to the usual fermion Yukawa couplings 
and the additional vertex
\begin{equation}
  {\cal L} = 2 \pi R y_{t} 
  (F_{Q}^{*} \phi_{U} \phi_{H} + F_{U}^{*} \phi_{Q} \phi_{H}).
\end{equation}

\begin{figure}
  \centerline{
  \psfig{file=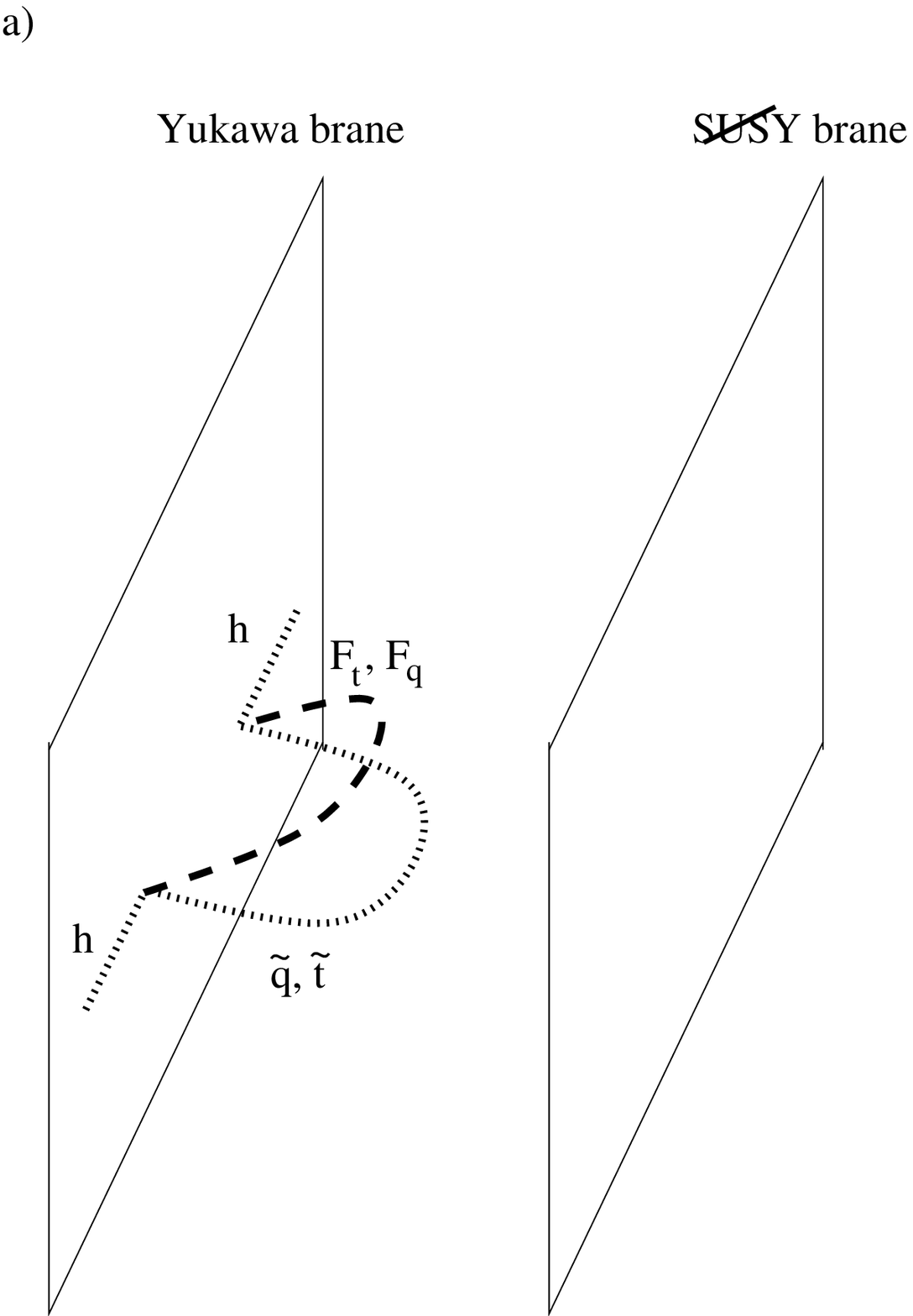,width=0.3\textwidth}\hskip 0.25in
  \psfig{file=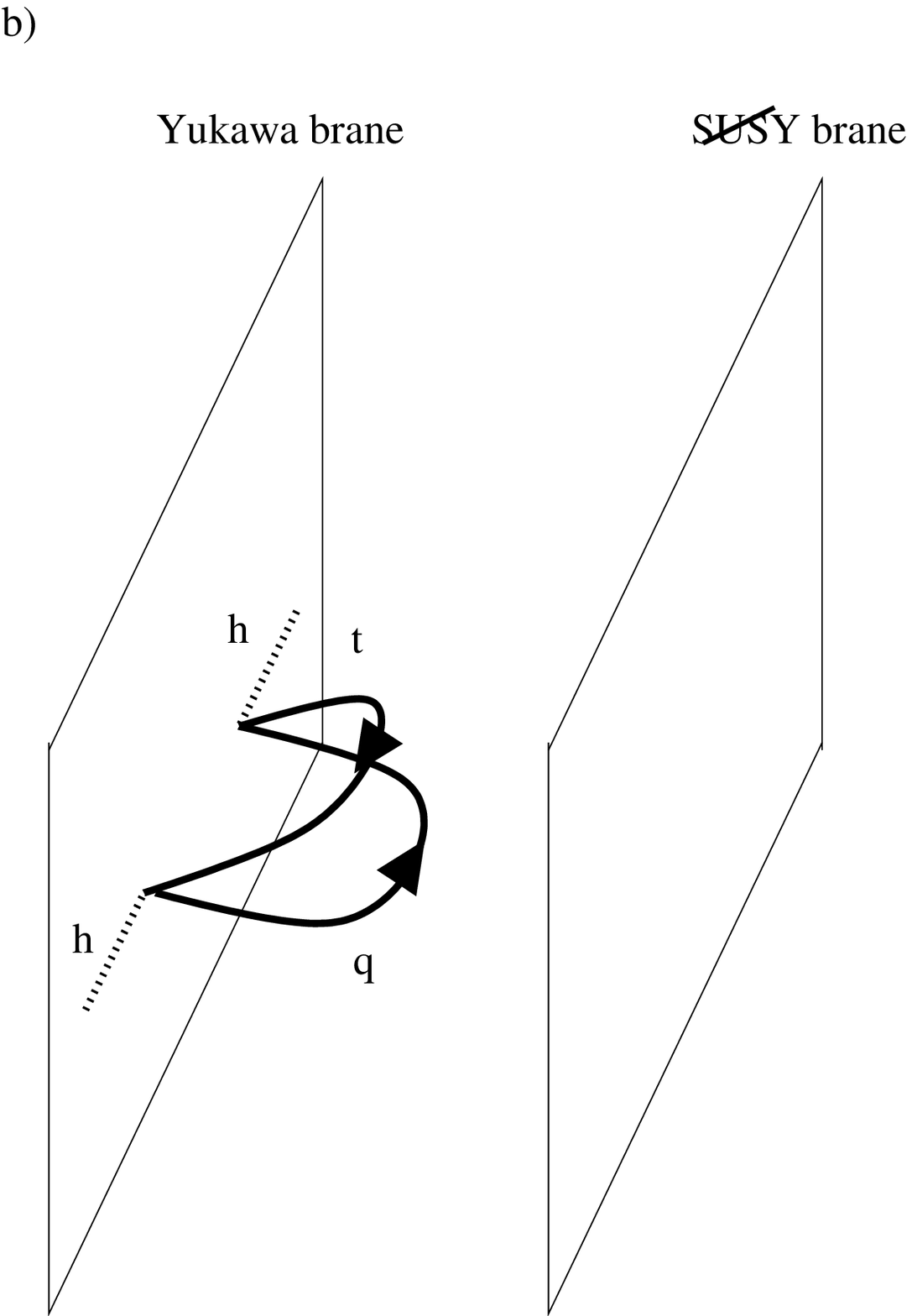,width=0.3\textwidth}}
\caption{Bosonic (a) and fermionic (b) contributions to the 
Higgs soft mass.}
\label{fig:5ddiags}
\end{figure}

The diagrams which contribute to the Higgs soft mass 
are given in Fig.~\ref{fig:5ddiags}. The relevant quantities are 
$\tilde G(k_{4},0)$, or the amplitude to propagate from the 
Yukawa brane back to the Yukawa brane.  We can derive these quantities 
by setting $y=0$ in the already obtained general propagators.
In the scalar case, for instance, this is given by
\begin{equation}
  \tilde G_{\phi}(k_{4},0) = \frac{1}{2 k_{4}}
  \left( \frac{2 k_4 \cosh[k_4 \pi R] + m \sinh[k_4 \pi R]}
  {2 k_4 \sinh[k_4 \pi R] + m \cosh[k_4 \pi R]} \right),
\label{eq:genprop-y0}
\end{equation}
from Eq.~(\ref{eq:genprop}).  In the opaque limit 
$m \rightarrow \infty$, it is further simplified as
\begin{equation}
  \tilde G_{\phi}(k_{4},0) = \frac{\tanh[k_{4} \pi R]}{2 k_{4}}.
\end{equation}

Now we calculate the Higgs-boson mass in the $m \rightarrow \infty$ 
limit.  The bosonic amplitude from Fig.~\ref{fig:5ddiags}(a) is 
given by
\begin{equation}
  m^{2}_{\rm boson} = 2 N_c \int \frac{d^{4}k_{4}}{(2 \pi)^{4}}
  (2 \pi R y_{t})^{2} \frac{\tanh[k_{4}\pi R]}{2 k_{4}}
  \frac{k_{4} \coth[k_{4} \pi R]}{2}, 
\end{equation}
and the fermionic one from Fig.~\ref{fig:5ddiags}(b) by
\begin{equation}
  m^{2}_{\rm fermion} = - N_c \int \frac{d^{4}k_{4}}{(2 \pi)^{4}}
  (2 \pi R y_{t})^{2} {\rm Tr}
  \left[\frac{\not k_{4} \coth[k_{4} \pi R]}{2 k_{4}}
  \frac{(1-\gamma_{5})}{2}
  \frac{\not k_{4} \coth[k_{4} \pi R]}{2 k_{4}}
  \frac{(1+\gamma_{5})}{2}\right].
\end{equation}
These are combined to give a total amplitude
\begin{equation}
  m^{2}_{\rm tot} = 2 N_c \int \frac{d^{4}k_{4}}{(2 \pi)^{4}}
  (2 \pi R y_{t})^{2} \frac{\coth[k_{4} \pi R]}{4} 
  \left( \tanh[k_{4} \pi R] - \coth[k_{4} \pi R] \right).
\end{equation}
We can rewrite this as
\begin{equation}
  m^{2}_{\rm tot} = - \frac{N_c\, y_{t}^{2}}{4 R^{2}}
  \int_0^{\infty}dx \frac{x^{3}}{\sinh^{2}[\pi x]} = 
  - \frac{3\, \zeta(3)}{8 \pi^{4}} \frac{N_c\, y_{t}^{2}}{R^{2}},
\end{equation}
which precisely reproduces the result obtained in the previous 
subsection.

\begin{figure}
  \centerline{
  \psfig{file=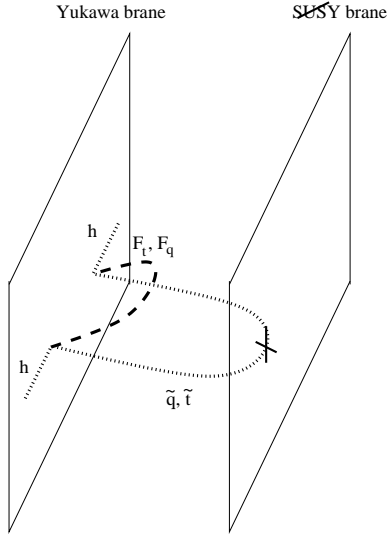,width=0.3\textwidth}}
\caption{The difference between bosonic and fermionic loops leaves 
just the reflected bosonic piece contributing to the Higgs soft mass.}
\label{fig:5dreflect}
\end{figure}

It may still not be completely transparent that it is the point 
splitting between Yukawa and supersymmetry-breaking branes 
that is responsible for this finite result. We have one more avenue to 
examine this, however. In the bosonic loop, the supersymmetry breaking 
is manifested in the presence of a ``reflected'' piece, which we 
illustrate in Fig.~\ref{fig:5dreflect}; the rest should 
all be cancelled from the fermionic loop. In lieu of this, we can forget 
about the fermion diagram entirely and obtain the correct result just 
by considering the reflected piece of the bosonic loop.

If we evaluate the scalar propagator of Eq.~(\ref{eq:genprop-y0}) 
in the $m\rightarrow 0$ limit, it is just 
\begin{equation}
  \tilde G_{\phi}(k_{4},0)=\frac{1}{2 k_{4}} \coth[k_{4} \pi R].
\end{equation}
If we subtracted this from the scalar propagator, we are left with 
just the reflected piece.  Then, if we calculate {\it only} the bosonic 
loop using this subtracted propagator, we find the same result as 
when we calculate {\it both} bosonic and fermionic loops using 
non-subtracted propagators.  Thus, we can see that only the reflected 
pieces are contributing to the Higgs-boson mass, and it is the point 
splitting which renders the contribution ultraviolet finite.

This is exciting, because it is {\em not} the presence of weak-scale 
supersymmetry that protects the Higgs mass, at least not in a 
conventional sense.  In the 4d picture, not only do we have many top-stop 
contributions to the mass, but also the fermion zero mode has a 
coupling differing by a factor of $\sqrt{2}$ from the scalar --- 
effectively {\em hard} supersymmetry breaking! 
It is for these reasons that it is imperative to understand the 
situation with the fifth dimension explicit.  Although the KK 
formalism more easily lends itself to calculations of effective 
potentials, the most important features of the model are transparent 
in mixed position-momentum space.

\section{Model with ${\bf Z}_{2,R}$ Scherk-Schwarz Mechanism}
\label{section:Z2R-SS}

\subsection{Setup}
\label{subsection:SSsetup}

The second model we consider is based on the Scherk-Schwarz mechanism 
\cite{Scherk-Schwarz} of supersymmetry breaking.
If the theory possesses a global symmetry, we can use it to modify 
the boundary conditions for various fields.
Specifically, the boundary condition for a field $\varphi$ is given by
\begin{equation}
  \varphi(y + 2\pi R) = {\cal S} \varphi(y),
\end{equation}
where ${\cal S}$ is a generator of the global symmetry.
If this symmetry is an $R$ symmetry, the bosonic and fermionic components 
in the same supermultiplet have different boundary conditions.
From the 4d point of view, this results in different masses for the bosonic 
and fermionic KK modes and thus breaks supersymmetry.

The question then is what $R$ symmetry a given supersymmetric 
theory possesses.  In the present framework we have brane Yukawa 
interactions given in Eq.~(\ref{eq:Yukawa-int}), so that it must be 
a symmetry respected by these interactions.  We take the simplest 
such possibility, $R$ parity, which is also 
anomaly free with respect to the standard-model gauge interactions.
Under $R$ parity, various superfields transform as
\begin{eqnarray}
\begin{array}{ll}
  X(x, y, \theta)      \rightarrow  -X(x, y, -\theta), \qquad &
  X^c(x, y, \theta)    \rightarrow  -X^c(x, y, -\theta),
\nonumber\\
  H(x, y, \theta)      \rightarrow   H(x, y, -\theta), \qquad &
  H^c(x, y, \theta)    \rightarrow   H^c(x, y, -\theta),
\\
  V(x, y, \theta)      \rightarrow   V(x, y, -\theta), \qquad &
  \Sigma(x, y, \theta) \rightarrow   \Sigma(x, y, -\theta),
\nonumber
\end{array}
\end{eqnarray}
where $X$ and $H$ represent $Q, U, D, L, E$ and $H_u, H_d$, respectively.

In this $Z_{2,R}$ Scherk-Schwarz model, the Higgs fields can be either 
bulk or brane fields.  If the Higgs supermultiplets live in the bulk, 
all the fields in the model have KK towers.  Then, since zero modes 
are contained only in the component fields which are even under both 
$Z_2$ orbifolding and the $R$ parity, the zero-mode sector of the model 
is the two Higgs-doublet standard model.
On the other hand, if the Higgs chiral multiplets are localized on the 
$y=0$ brane, they do not have KK excitation and neither their fermionic 
nor bosonic components obtain masses from the boundary condition.
In the next subsection, we analyze wavefunctions of the fields and 
derive the KK mass spectrum of the model at the tree level.  
The one-loop Higgs mass-squared are also calculated.

\subsection{Particle spectrum and radiatively induced $m_{H_u}^2$}
\label{subsection:SSspectrum}

The bulk fields are classified into 4 types according to their 
transformation properties under the $Z_2$ orbifolding and the $R$ 
parity.  Each class has a mode expansion of the following form:\footnote{
This mode expansion is equivalent to that obtained by compactifying 
the extra dimension on the $S^1/(Z_2 \times Z_2')$ orbifold 
\cite{Barbieri:2000vh}.}
\begin{eqnarray}
&& (+,+): \;\;\; \cos{n\,y \over R} \\
&& (+,-): \;\;\; \cos{(n+1/2)\,y \over R} \\
&& (-,+): \;\;\; \sin{(n+1)\,y \over R} \\
&& (-,-): \;\;\; \sin{(n+1/2)\,y \over R}
\label{eq:z2z2modes}
\end{eqnarray}
with $n=0,1,2,...$.  Here, the first and the second signs correspond 
to the quantum numbers under the $Z_2$ orbifolding and the $R$ parity, 
respectively.  Only fields with (+,+) assignment contain a zero mode.

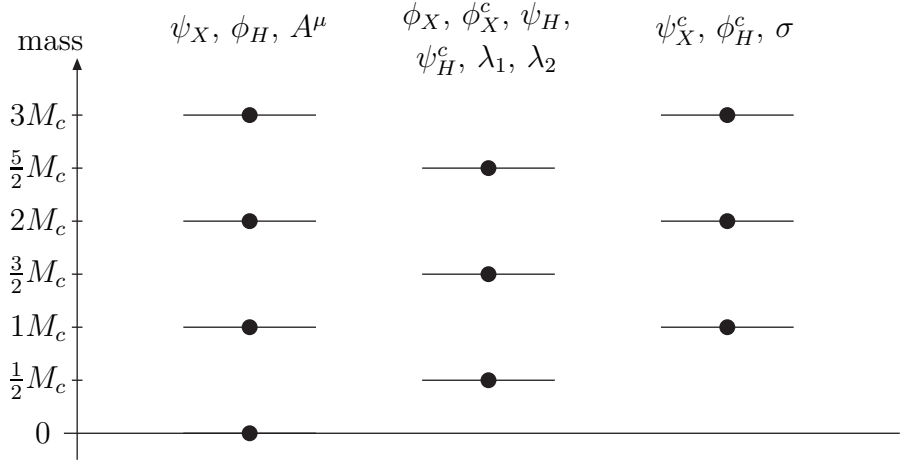
\begin{figure}
\begin{center} 
\begin{picture}(350,190)(-10,-25)
  \Line(5,0)(320,0)
  \LongArrow(10,-10)(10,140)
  \Text(0,145)[b]{mass}
  \Text(0,0)[r]{$0$}
  \Line(8,20)(12,20)    \Text(6,20)[r]{${1\over 2}M_c$}
  \Line(8,40)(12,40)    \Text(6,40)[r]{$1 M_c$}
  \Line(8,60)(12,60)    \Text(6,60)[r]{${3 \over 2}M_c$}
  \Line(8,80)(12,80)    \Text(6,80)[r]{$2 M_c$}
  \Line(8,100)(12,100)  \Text(6,100)[r]{${5 \over 2} M_c$}
  \Line(8,120)(12,120)  \Text(6,120)[r]{$3 M_c$}
  \Text(75,150)[b]{$\psi_X,\, \phi_H,\, A^{\mu}$}
  \Line(50,0)(100,0)      \Vertex(75,0){3}
  \Line(50,40)(100,40)    \Vertex(75,40){3}
  \Line(50,80)(100,80)    \Vertex(75,80){3}
  \Line(50,120)(100,120)  \Vertex(75,120){3}
  \Text(165,155)[b]{$\phi_X,\, \phi_X^c,\, \psi_H,$}
  \Text(165,147)[t]{$\psi_H^c,\, \lambda_1,\, \lambda_2$}
  \Line(140,20)(190,20)    \Vertex(165,20){3}
  \Line(140,60)(190,60)    \Vertex(165,60){3}
  \Line(140,100)(190,100)  \Vertex(165,100){3}
  \Text(255,150)[b]{$\psi_X^c,\, \phi_H^c,\, \sigma$}
  \Line(230,40)(280,40)    \Vertex(255,40){3}
  \Line(230,80)(280,80)    \Vertex(255,80){3}
  \Line(230,120)(280,120)  \Vertex(255,120){3}
\end{picture}
\caption{Mass spectrum for the lowest KK modes of $Z_{2,R}$ 
  Scherk-Schwarz model with the bulk Higgs fields.}
\label{fig:spectrum-2}
\end{center}
\end{figure}
The spectrum of states is summarized in Fig.~\ref{fig:spectrum-2} 
in the case where the Higgs fields live in the bulk.
In the figure, we can see that the zero-mode sector is the standard 
model with two Higgs doublets.  On the other hand, if two Higgs chiral 
multiplets are localized on the brane, they do not have KK excitations 
and all the component fields are massless at this level.  
In this case, the Higgsino mass should somehow be generated to make 
the model phenomenologically viable.  We will see how this can be 
done naturally in section \ref{section:Higgs} where the Higgs sector 
of the model is discussed in detail.

In either case of the bulk or brane Higgs fields, the KK mass spectrum 
for the quark multiplets is the $(r^{\phi}, r^F) = (1/2, 1/2)$ case 
of section \ref{section:rad-cor}.  Thus, using Eq.~(\ref{mH-2-1}), 
we obtain a radiatively induced up-type Higgs-boson mass
\begin{equation}
  m_{\phi_{H_u}}^2 = - {21\, \zeta(3) \over 32 \pi^4} N_c\, y_t^2 M_c^2,
\label{eq:HG-2}
\end{equation}
which is finite and one-loop suppressed compared with $M_c$, but has 
a slightly different coefficient from that in the model of section 
\ref{section:localized} due to the different KK spectrum for $\phi_X^c$.

\subsection{5d interpretation}

Just as in the model with supersymmetry breaking localized on a brane, 
the ultraviolet finiteness of this model is intimately related to its 
five dimensional features. However, there is nothing so intuitively 
simple as a supersymmetry-breaking ``reflected'' piece as before. It 
is the difference in boundary conditions that breaks supersymmetry, 
rather than an $F$-component vacuum expectation value.  
The supersymmetry breaking will appear in our calculations as a 
different sign picked up when propagating multiple times around the 
fifth dimension.

We begin by considering the scalar propagator. It gives a minus sign 
when we go around the circle $y \rightarrow y + 2\pi R$.
We represent this by having alternating signs in our image sources, 
giving a propagator
\begin{equation}
  \tilde G_{\phi}(k_{4},y) 
  = \sum_{n=-\infty}^{\infty} \frac{1}{2 k_{4}} (-1)^{n} 
    e^{-k_{4} |y - 2 \pi n R|} 
  \stackrel{y=0}{\longrightarrow} \frac{1}{2 k_{4}} \tanh[k_{4}\pi R].
\label{eq:ssscalarprop}
\end{equation}
We can do the same thing for the $F$-component propagator and obtain
\begin{equation}
  \tilde G_{F}(k_{4},y) 
  = \sum_{n=-\infty}^{\infty} \frac{k_{4}}{2} (-1)^{n}
    e^{-k_{4} |y - 2 \pi n R|}
  \stackrel{y=0}{\longrightarrow} \frac{k_{4}}{2} \tanh[k_{4}\pi R],
\label{eq:ssfprop}
\end{equation}
and likewise for the fermion
\begin{equation}
  \tilde G_{\psi}(k_{4},y) 
  = \sum_{n=-\infty}^{\infty} \frac{\not k_{4}}{2 k_{4}}
    e^{-k_{4} |y - 2 \pi n R|}
  \stackrel{y=0}{\longrightarrow} \frac{\not k_{4}}{2 k_{4}} 
    \coth[k_{4}\pi R].
\label{eq:ssfermionprop}
\end{equation}

We can now calculate the amplitudes for the Higgs-boson mass
in this model as
\begin{eqnarray}
  m^{2}_{\rm boson} 
  &=& 2 N_{c} (2\pi R y_{t})^{2} 
    \int \frac{d^{4}k}{(2 \pi)^{4}}
    \frac{\tanh[k_{4} \pi R]}{2 k_{4}} 
    \frac{k_{4} \tanh[k_{4} \pi R]}{2} 
\nonumber\\
  &=& \frac{N_{c}\, y_{t}^{2}}{4 R^{2}} 
    \int_0^{\infty}dx\, x^{3} \tanh^{2}[\pi x],
\end{eqnarray}
and 
\begin{eqnarray}
  m^{2}_{\rm fermion} 
  &=& - N_{c} (2 \pi R y_{t})^{2} 
    \int \frac{d^{4}k}{(2 \pi)^{4}} 
    {\rm Tr}\left[\frac{(1-\gamma_{5})}{2}
    \frac{\not k_{4} \coth[k_{4} \pi R]}{2 k_{4}}
    \frac{(1+\gamma_{5})}{2}
    \frac{\not k_{4} \coth[k_{4} \pi R]}{2 k_{4}}\right]
\nonumber\\
 &=& - \frac{N_{c}\, y_{t}^{2}}{4 R^{2}}
    \int_0^{\infty}dx\, x^{3} \coth^{2}[\pi x],
\end{eqnarray}
which gives a total contribution to the Higgs-boson mass
\begin{equation}
  m^{2}_{\rm tot} = 
    - \frac{N_{c}\, y_{t}^{2}}{4 R^{2}}
    \int_0^{\infty}dx\, x^{3} 
    \left( \coth^{2}[\pi x]-\tanh^{2}[\pi x] \right) 
  = -\frac{21\, \zeta(3)}{32 \pi^{4}} 
    \frac{N_{c}\, y_{t}^{2}}{R^{2}},
\end{equation}
again, reproducing our result from the KK calculation.
Unfortunately, there is no parameter that we can continuously 
vary to return to the supersymmetric case, and the trick we noted 
previously of only using the reflected piece will not work here. 
However, we can still understand the finite result by analyzing 
the propagators in Eqs.~(\ref{eq:ssscalarprop} -- \ref{eq:ssfermionprop}). 
If we keep only the $n=0$ piece in all of the expansions, we would 
find that the different pieces cancel. It is only the winding modes 
that are sensitive to the different boundary conditions, and thus 
only these modes can contribute to the total Higgs soft mass. 
At four momenta greater than $(\pi R)^{-1}$, the Higgs does not 
``see'' that the dimension is compact and does not notice the 
presence of the winding modes, thus the final result is again 
ultraviolet finite.

\section{The Higgs Sector}
\label{section:Higgs}

In this section we demonstrate how our mechanisms for generating 
a finite, negative $m_{H_u}^2$ from the bulk can be utilized in 
Higgs sectors that give realistic EWSB.

\subsection{The minimal sector}

Consider first the supersymmetry-breaking brane model of section 
\ref{section:localized}, in which the Higgs doublets are localized on 
the Yukawa brane.  In this case one can simply add the superpotential term
\begin{equation}
  W_{\rm Higgs}=\mu H_u H_d, 
\label{eq:branemu}
\end{equation}
and generate $B \mu$ term (an analytic supersymmetry-breaking mass for 
the Higgs doublets) radiatively from the $\mu$ term and gaugino masses.
Because $\mu$ and $m_{\lambda}$ arise on different branes, the loop
integral is regulated by the compactification scale in much the
same way as the radiative corrections to the Higgs-boson mass, giving
\begin{equation}
  B \mu \sim {1 \over 16 \pi^2} \mu M_c.
\label{eq:bmu}
\end{equation}
The standard relations for the Higgs sector of the minimal supersymmetric 
standard model (MSSM) apply; in particular, we have
\begin{equation}
  \sin 2 \beta ={2 B \mu \over 2 \mu^2+m_{H_u}^2+m_{H_d}^2}.
\label{eq:beta}
\end{equation}
Here $m_{H_u}^2$ is given by Eq.~(\ref{eq:HG-1}), and we can neglect
$m_{H_d}^2$ for simplicity.  For successful EWSB, 
we need $|\mu|^2 \sim |m_{H_u}^2| \sim M_c^2/\pi^4$.  
Because $B \mu$ arises from $\mu$ as a loop effect, 
one then expects from Eq.~(\ref{eq:beta}) that $\tan\beta$ will be
somewhat large, $\tan\beta \sim O(10)$.\footnote{Interestingly, in 
the limit of very strong supersymmetry breaking, $F_Z\gg M_{*}^2$, 
the gaugino wavefunction is repelled from the supersymmetry-breaking 
brane and does not pick up a Majorana mass, so that $B \mu$ is not 
generated.  If instead $F_Z \sim  M_{*}^2$, then Eq.~(\ref{eq:bmu}) 
holds.}

It is unavoidable in this model that one must take $\mu$ to be
suppressed relative to the fundamental scale, 
$\mu \sim M_c/\pi^2$.  On the other hand,
once one accepts this suppression, acceptable symmetry breaking
can be achieved without a severe fine tuning of $\mu$ relative to
$m_{H_u}^2$ even for $M_c \simeq 3$ TeV.

From the viewpoint of experiment, this theory would appear much like the 
MSSM, but with heavy matter and gauge superpartners. The three free 
parameters that determine the Higgs sector
characterize: the matter superpartners degenerate at $M_c/2$, 
the mass of the lightest mode of the neutral wino at a somewhat lower 
scale $m_\lambda$, and the Higgsinos nearly degenerate at a significantly 
lower scale $\mu$. All other observables, such as the masses of the 
charged and neutral Higgs bosons and $\tan\beta$ are predicted 
in terms of these parameters, providing further tests of the theory.

The Yukawa-brane superpotential of Eq.~(\ref{eq:branemu}) is not
sufficient for the Scherk-Schwarz model of section \ref{section:Z2R-SS}, 
because the gauginos do not have the Majorana masses required
for the loop diagram that generates $B \mu$ in the supersymmetry-breaking
brane model.  We then have the usual problems that
a vanishing $B \mu$ leads to in the MSSM:  the scalar potential possesses
a Peccei-Quinn symmetry, and either $H_u=H_d$ is a runaway
direction or at least one of $H_u$ and $H_d$ is stabilized at the
origin.  Regardless of whether or not the Higgs fields propagate in the bulk,
these problems can be resolved by adding a
non-renormalizable $(H_u H_d)^2$ term to the Yukawa-brane superpotential.
Starting with
\begin{equation}
  W_{\rm Higgs}=\mu H_u H_d +{\lambda \over M_{*}} (H_u H_d)^2,
\label{eq:sp2}
\end{equation}
we obtain the scalar potential 
\begin{equation}
V=m_{H_u}^2\left| H_u^0\right| ^2  +  \left| \mu H_u^0
  -2{\lambda \over M_{*}} H_d^0
  {H_u^0}^2 \right|^2 +\left| \mu 
H_d^0 -2{\lambda \over M_{*}} H_u^0 {H_d^0}^2 \right|^2 +{g^2 + {g'}^2 \over
  8}\left( \left| H_u^0\right| ^2-\left| H_d^0\right| ^2 \right)^2.
\label{eq:p1}
\end{equation}  
The terms in the potential proportional to $|\lambda|^2$ remove any
possibility of runaway behavior, and the cross terms in the $|F|^2$
pieces of the potential generate an effective $B\mu$ term:
\begin{equation}
  V\supset-2 {\lambda \mu^{*} v^2\over M_{*}} H_u^0 H_d^0 + {\rm h.c}.
\end{equation} 
We find that acceptable symmetry breaking occurs for $\mu^2 \sim
|m_{H_u}|^2$ and $\lambda \simeq (0.1 \sim 0.3) M_{*}/M_c.$  To obtain a 
large enough mass for the lightest Higgs boson and $M_c \sim 1$ TeV, 
$\mu$ must be chosen to be within roughly 30\% of $|m_{\phi_{H_u}}|$.  
For this model we again find that $\tan\beta$ tends to be large.  

If one employs the superpotential of Eq.~(\ref{eq:sp2}) with the Higgs
doublets localized on the Yukawa brane, then one must suppress $\mu$
by hand as in the case of the supersymmetry-breaking brane model.  
If, on the other hand, the Higgs doublets propagate in the bulk, then 
$\mu$ is volume suppressed and scales as $M_c/\pi$.  
In this case, one also expects $\lambda$ to be suppressed, by a factor
of $(M_c/(\pi M_{*}))^2$.  To obtain the correct size for
$\lambda$ for acceptable symmetry breaking, the coupling of
the 5d fields must then be rather large, $\sim 10/M_{*}^3$.

For large $\tan\beta$, the spectrum of Higgs masses in this model is 
quite insensitive to the parameter $\lambda$, which always appears in
front of at least one factor of $H_d$ in the potential. One cannot
take $\tan\beta$ to be arbitrarily large because at some point it becomes
inconsistent: radiative corrections induced by a large $\lambda_b$ drive
a large vacuum expectation value for $H_d$.  But for a range of 
moderately large $\tan\beta$ ($\sim 20$), the Higgs masses are 
approximately equal to their values in the $\lambda \rightarrow 0$ limit. 
One combination of $\mu$ and $M_c$ is fixed by requiring that 
minimization of the potential yields the correct value for 
the Fermi constant.  Thus, specifying either $M_c$ or $\mu$ determines 
each of the charged and neutral Higgs masses -- as well as the masses 
of the squarks and sleptons -- for this range of $\tan\beta$.

In this regime, $H_u^0$ and $H_d^0$ are approximate mass eigenstates.
The mass of the down-type Higgs is 
\begin{equation}
  m_{H_d}^2 \simeq |\mu|^2-\frac{m_Z^2}{2}.
\end{equation}
For an accurate determination of the up-type Higgs boson one needs 
the one-loop effective potential $V(H_u)$, which has been calculated 
for an identical spectrum of quark and squark modes in 
Ref.~\cite{Barbieri:2000vh}. (We need only the trivial replacement 
$R \rightarrow 2R$.) Using this effective potential, one can show that
for $\mu^2$ large compared to $m_Z^2/2$,
the requirement that the supersymmetric and radiatively generated
contributions to $m_{H_u}^2$ cancel at leading order gives a simple
linear relation between $\mu$ and $M_c$:
\begin{equation}
  M_c\simeq {2\pi^2 \over 3}{v \over m_t}\mu.
\end{equation}
By ignoring $\lambda$ and using the effective potential of 
Ref.~\cite{Barbieri:2000vh}, we calculate Higgs masses that can be 
trusted at roughly the 10\% level, with uncertainties arising from
$1/\tan\beta$ effects, neglected gauge loop contributions, two-loop
effects, and uncertainty in the mass of the top quark, for instance.

\begin{figure}
\vspace{-3.25in}
\psfig{file=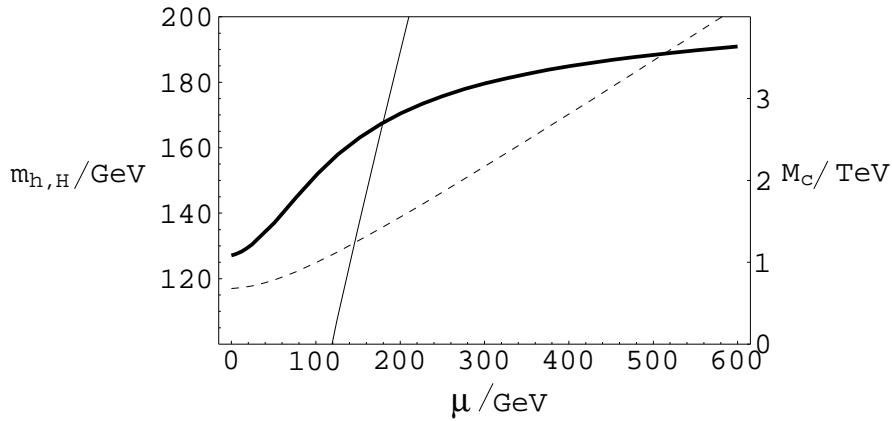,width=\textwidth}
\vspace{-3.25in}
\caption{The compactification scale and masses of the up- and down-type 
  Higgs scalars, plotted as functions of $\mu$ in the $\lambda\rightarrow 0$ 
  limit of our $Z_{2,R}$ Scherk-Schwarz model.  The darker
  solid line is $m_{H_u}$, the lighter solid line is $m_{H_d}$, and
  the dashed line is $M_c$. We extend the plot to $\mu=0$ simply to
  show that the prediction $m_h=127$ GeV of Ref.~\cite{Barbieri:2000vh} 
  is recovered. (In Ref.~\cite{Barbieri:2000vh}, only a single Higgs 
  doublet exists in the low energy theory.)}
\label{fig:mudep}
\end{figure}

We plot $m_{H_u}$, $m_{H_d}$, and $M_c$ as a function of $\mu$ in the
$\lambda \rightarrow 0$ limit in Fig.~\ref{fig:mudep}.  The plot 
illustrates that the MSSM bound on the mass of the lightest Higgs boson, 
$m_h \lsim 135$ GeV, is easily evaded in the present model, but that 
the mass range for the lightest Higgs still lies largely within 
that preferred by precision electroweak data \cite{Higgs-LEP}. 
Note that we evade the MSSM bound because the low energy theory is not
the MSSM.  For instance, at the lightest KK level we have both
squarks and conjugate squarks, and these fields' couplings to the
Higgs zero mode are enhanced relative to those of the zero mode quarks
by factors of $\sqrt{2}$.
Although the quantitative results of the plot apply only to the 
Scherk-Schwarz model of section \ref{section:Z2R-SS}, a similar situation 
exists for the supersymmetry-breaking brane model with radiatively
generated $B \mu$: for moderately large $\tan\beta$, the
Higgs spectrum and the spectra of fermion and sfermion KK modes are all
determined by a single additional free parameter, which may be taken to be
either $\mu$ or $M_c$.

\subsection{Generating $\mu$: the next to minimal sector}

So far we have regarded $\mu$ as an input parameter, but in both the 
supersymmetry-breaking brane and Scherk-Schwarz models, $\mu$ could 
instead be generated dynamically as in the next to MSSM (NMSSM).
Consider the case where the Higgses are brane fields, and take the 
superpotential
\begin{equation}
  W_{\rm Higgs} = \lambda_H S H_{u}H_{d} + \lambda_B S B \bar{B}
  +{\lambda_S \over 3}S^3. 
\end{equation}
Here we take $S$, $B$ and $\bar{B}$ to be standard-model singlets, with 
$S$ a brane field and $B, \bar{B}$ part of a bulk hypermultiplet.  This
superpotential is justified by a $Z_3$ symmetry under which each
superfield has charge $+1$.  In the Scherk-Schwarz model, 
the transformation properties of $B$ and $\bar{B}$ under
$Z_{2,R}$ are the same as those of $Q$,$U$,$D$,$L$,$E$.  

The $B$ and $\bar{B}$ fields drive the mass-squared of the $S$ scalar 
negative just as the top and stop fields drive $m_{H_{u}}^2$ negative. 
The formula for $m_{S}^2$ is given by either Eq.~(\ref{eq:HG-1}) or 
Eq.~(\ref{eq:HG-2}), with $y_t$ replaced by $\lambda_B$, and $N_c$ 
interpreted as the multiplicity of $B$ and $\bar{B}$ states. 
The $S$ scalar is forced to acquire a vacuum expectation value, 
which in turn induces an $F_S$ expectation value through the 
$S^3$ term in the superpotential.  These vacuum expectation values give 
rise to an effective $\mu$ and an effective $B \mu$, respectively.  
They tend to give $B^2 \sim \mu^2 \sim |m_{H_{u}}|^2$, provided 
$\lambda_B$ and the multiplicity of $B$ and $\bar{B}$ states are 
chosen so that $m_{S}^2 \sim m_{H_{u}}^2$.

To obtain an $S$ vacuum expectation value, $\lambda_H$ must be 
somewhat small ($\lsim 1/3$), so that the positive mass-squared 
coming from $\lambda_H^2 (|H_u|^2+|H_d|^2)|S|^2$ in the potential does not 
overwhelm the negative contribution from $B$ loops.  To obtain large 
enough Higgsino mass, one must then choose $\lambda_S$ to be somewhat 
small as well, but not so small that the potential becomes stable 
about $\langle H_u \rangle = \langle H_d \rangle =0$.  We find that 
this requirement can be met if $\lambda_S$ is chosen with roughly 
5\% precision.  Given satisfactory parameters, one finds 
$M_c \simeq (1 \sim 3)$ TeV. 

One could alternatively place both the Higgs and $S$ fields in the
bulk for the Scherk-Schwarz case (but not for the supersymmetry-breaking 
brane model).  In this case, couplings of the 5d fields of order
$10/M_{*}^{3/2}$  must be chosen to compensate for the volume suppressions 
of the fields' couplings to the Yukawa brane.

\section{Phenomenology}
\label{section:pheno}

In this section, we discuss how the compactification and the cutoff 
scales are determined and give an idea about various scales appearing 
in the theories.  We also briefly discuss some phenomenological issues 
in the models presented in sections \ref{section:localized} and 
\ref{section:Z2R-SS}.

\subsection{Compactification and cutoff scales}

We first consider the compactification scale.
As we have seen in section \ref{section:rad-cor}, the compactification 
scale $M_c$ is related to the soft supersymmetry-breaking mass for the 
up-type Higgs boson by a one-loop factor:
\begin{eqnarray}
  m_{H_u}^2 \simeq - \frac{1}{\pi^4} M_c^2.
\end{eqnarray}
In the theories with a Higgs sector of the MSSM type, as in the first example 
of the previous section, the condition for EWSB gives 
\begin{eqnarray}
  \frac{m_Z^2}{2} \simeq - m_{H_u}^2 - |\mu|^2.
\label{eq:EWSB-cond}
\end{eqnarray}
Combining these two equations, we find natural size of $M_c$ to be 
$(\pi^2/\sqrt{2}) m_Z \sim 600~{\rm GeV}$.
However, this value has an order-one ambiguity coming from the presence 
of the second term in Eq.~(\ref{eq:EWSB-cond}) and various numerical 
factors omitted in the above equations.
Thus, here we take $M_c \simeq 1~{\rm TeV}$ as a representative value.
This value is easily realized by choosing parameters in the model.
It is interesting to note that having $M_c \simeq 1~{\rm TeV}$ 
($500~{\rm GeV}$ squarks and sleptons) requires only a factor of $3$ 
cancellation between two terms in Eq.~(\ref{eq:EWSB-cond}).
This is in contrast to the usual 4d supersymmetric theories, where we 
need an order of magnitude cancellation to obtain corresponding sfermion 
masses.  The crucial difference between the two theories is that in our 
case the Higgs soft mass is generated only by a one-loop diagram 
proportional to 
squark masses, while in the case of usual 4d theories the Higgs and the 
sfermion masses are generated at the same loop orders.  As a consequence, 
the squark and slepton masses are naturally larger than the soft masses 
for the Higgs boson in our theories.

In the NMSSM-type theories, like the last example in the previous 
section, the EWSB condition is somewhat different due to the presence 
of an extra quartic coupling for the Higgs bosons.
The $\mu$ parameter is replaced by 
the vacuum expectation value of the singlet field $S$.
The situation, however, is similar, with $M_c \simeq 1~{\rm TeV}$ 
obtained by choosing parameters in the model to give 
the required cancellation. 

With the above value $M_c \simeq 1~{\rm TeV}$, there is no stringent 
experimental bound coming from direct production of the KK gauge 
bosons \cite{Barbieri:2000vh, Appelquist:2000nn}.  
This is because we have put the quark and lepton multiplets in the bulk 
so that there is no interactions between the zero-mode fermions and 
the excited modes of the gauge bosons, 
$\bar{\psi}_0 \gamma_\mu A^\mu_k \psi_0$, to the leading order, 
due to momentum conservation in the extra dimension.
It also ensures that dangerous operators such as four-fermion operators 
and operators which causes mass mixing between the electroweak gauge 
bosons and their excited modes are not generated at the tree level.
At the loop level, however, we have electroweak observables such as 
$\rho$ parameter which are quadratically sensitive to the ultraviolet 
physics \cite{Barbieri:2000vh}.  These quantities are not reliably 
calculated in the effective field theory, but we can estimate the 
contributions from the lowest-lying top-stop KK towers.  If $M_c$ is 
higher than a few TeV, the contributions are smaller than the experimental 
upper bounds, since they scale as $1/M_c^2$.  If $M_c$ is lower, on the 
other hand, they have to be cancelled by other contributions coming 
from underlying physics at the cutoff scale.

We next consider the cutoff scale $M_*$.  Since higher-dimensional 
field theories are in general non-renormalizable theories, they must 
be regarded as cutoff theories.  Then, the upper bound on the cutoff 
scale comes from the strength of interactions in the low-energy 4d 
effective theory.  To see this explicitly, let us consider the top-Yukawa 
coupling in the brane Higgs case.  In 5d, the Yukawa coupling 
is written as $\delta(y) \int d^2\theta (f_t/M_*) Q_3 U_3 H_u$, where 
the dimensionless coupling $f_t$ is bounded as $f_t \lsim 6 \pi^2$ by 
the strong coupling analysis in higher dimensions \cite{Chacko:2000hg}.  
Since the 4d top-Yukawa coupling is given by $y_t = (f_t/2\pi)(M_c/M_*)$, 
we find that $y_t \sim 1$ gives an upper bound on the cutoff scale 
$M_* \lsim 3\pi M_c$.  The same can also be seen from the 4d point of view 
by making a KK decomposition.  At energy $E$, the loop expansion 
parameter is given by $(y_t^2/16\pi^2) N_{\rm KK}^2$ where 
$N_{\rm KK} \simeq 2E/M_c$ is the KK multiplicity.  Thus, in order for 
the theory to make sense, the expansion parameter should be smaller 
than 1, giving $M_* \lsim 2\pi M_c$.

In addition to the above effect, we also have power-law runnings of 
the couplings \cite{Dienes:1998vh}.  Since the top-Yukawa coupling is 
asymptotically non-free, this effect makes the bound on $M_*$ tighter.
Paying careful attention to the thresholds of the KK excitations, we 
finally find that $M_* \lsim 4 M_c$ and $M_* \lsim 2.5 M_c$ in the 
brane and bulk Higgs cases, respectively.  Therefore, we consider that 
our theories are cut off and embedded in some more fundamental theory 
such as string theory, or approach to some strongly-coupled ultraviolet 
fixed point, at these scales.  In the former case, the observed 
smallness of gravity may be understood by the presence of additional 
large extra dimensions in which only gravity propagates \cite{ADD}.

\subsection{Suppression of flavor violation from squark and slepton exchange}

Although we will not specify the physics which gives the observed 
quark and lepton mass matrices, we discuss 
some aspects of flavor physics in our theories.
In any extension of the standard model which introduces new physics 
at the TeV scale, the question of suppressing flavor changing neutral 
currents (FCNCs) must be addressed.  In our case, since the theories 
are cut off at the multi-TeV scale, there are two different sources 
for FCNC processes.  One comes from unknown ultraviolet physics, and is 
parameterized in the low energy theory by a set of higher dimensional
operators with coefficients suppressed by inverse powers of
$M_*$. Since $M_*$ is only a few TeV, the dimensionless couplings for
flavor-changing operators must be small.  A discussion of these operators 
and possible solutions with TeV cutoffs has been given in 
Ref.~\cite{Arkani-Hamed:2000yy}, and we will not attempt to address 
it here.

A second source of flavor violation is in the squark and slepton mass
matrices and the trilinear scalar interactions. Our theories provide
an explanation for the smallness of these contributions. In all our
theories, flavor symmetry breaking occurs only on the brane at $y =
0$, while supersymmetry breaking is not localized at this point.
This is the origin of the absence of the scalar trilinear
interactions, or $A$ terms. In the Scherk-Schwarz theory the scalar
masses are degenerate at the tree level. Non-degeneracies arise only from
radiative corrections involving the brane Yukawa matrices, and are
therefore safely under control. In the case of supersymmetry breaking
localized on a brane, the interactions which yield squark masses,
Eq.~(\ref{eq:ssbint}), may have large flavor violation, with couplings to 
one flavor of quark very different to that for another flavor. In the 
case that these couplings are all large, a dynamical near degeneracy
occurs. To leading order all squarks are degenerate with mass $M_c/2$,
independent of the size of these couplings: as long as the couplings
are large the masses are simply set by the geometry of the wavefunction. 
To next order, from Eq.~(\ref{eq:scalar-corr}), we find non-degeneracies 
\begin{equation}
  {m_i^2 - m_j^2 \over m^2} \simeq \left( {1 \over c_i} - {1 \over c_j}
  \right)  {M_*^4 \over F_Z^2} {M_c \over M_*}.
\label{eq:scalarnondeg}
\end{equation}
For strong enough supersymmetry breaking, or equivalently, large
enough couplings $c_{i,j}$, sufficient scalar degeneracy results,
even with 100\% differences between the couplings. For example, with
$F_Z = M_*^2$ and $c_{i,j}$ approaching strong coupling values of $24
\pi^3$, the degeneracy is much larger than needed to satisfy the 
$K_L$--$K_S$ mass difference constraint, especially as heavy squarks are
expected in these theories.

\subsection{Superpartner spectrum}

In this subsection, we discuss some aspects of the superpartner 
spectrum in our theories.  First, we ask what are the lowest states 
beyond those of the standard model.  In both theories of local and 
non-local supersymmetry breaking, we have superpartners of masses 
$M_c/2$.  However, the properties of these states are quite different 
in each case.

Let us begin with the squarks and sleptons.  In the case of localized 
supersymmetry breaking, we have only one set of superpartners as in 
the usual 4d supersymmetry.  On the other hand, in the Scherk-Schwarz 
case, we have two superpartners for every standard-model particle, 
reflecting underlying $N=2$ supersymmetry in the 5d theory.
This will give an unambiguous distinction between these two theories.
However, it is worth stressing that there is no energy interval where 
physics is described by a 4d supersymmetric theory even in the localized 
supersymmetry-breaking case.  The Yukawa couplings of the quarks and 
squarks are not related in the usual 4d supersymmetric way; there is a 
$\sqrt{2}$ factor coming from the difference of normalizations between 
the scalars and the zero-mode fermions.

As for the gauginos, we have two gauginos for each gauge boson 
in both models.  In the model with localized supersymmetry breaking, 
there are two Majorana gauginos whose masses could differ from $M_c/2$ 
by an order one constant, unless the coupling $c_W$ is very large.
While in the Scherk-Schwarz model, the two gauginos form one Dirac fermion 
whose mass is almost precisely $M_c/2$.  This point could also be used 
to discriminate between the models in future experiments.

Finally, we consider the nature of the LSP in our theories.
In the model of localized supersymmetry breaking, supersymmetry is 
broken by an $F$-component vacuum expectation value of the field $Z$.
Since we are considering $F_Z \simeq M_*^2$, we expect the gravitino 
to be very light of mass given by $m_{3/2} \simeq F_Z/M_{\rm pl} 
\simeq M_*^2/M_{\rm pl}$.  Therefore, the gravitino is the LSP in this 
localized supersymmetry breaking model.  With $M_* \simeq 
(3 \sim 5)~{\rm TeV}$, we find the gravitino mass $m_{3/2} \simeq 
(2 \sim 6)~{\rm meV}$.  The next to the LSP (NLSP) depends on the 
details of the model, but since the Higgsino is localized on the Yukawa 
brane and does not feel supersymmetry breaking directly, it is likely 
to be the NLSP.  Then, the Higgsino could decay into the Higgs boson 
and the gravitino inside the detector. At hadron colliders, if the
gluino is light enough it will be copiously produced and will decay
via $\tilde{g} \rightarrow \bar{q} q \tilde{h}$, followed by
$\tilde{h} \rightarrow h \tilde{G}$. This provides the dominant
mechanism for Higgs production, which therefore appear in pairs in
events with jets and large missing transverse energy.

In the $Z_{2,R}$ Scherk-Schwarz model, the gravitino obtains a mass 
of $M_c/2$ from the boundary condition and is not the LSP.
Then, there are two different possibilities for the LSP.  If the Higgs 
fields are the brane fields, the Higgsino is likely to be the LSP since 
it does not acquire $M_c/2$ mass from the boundary condition.
Thus, in this case, there are four fermions close in mass: two neutral 
and two charged ones.  On the other hand, if the Higgs fields are the 
bulk fields, all the superpartners acquire $M_c/2$ mass from the 
Scherk-Schwarz boundary condition.  This degeneracy of superpartner 
masses is lifted by the Higgs vacuum expectation values.
We then find the LSPs to be two top squarks of mass $M_c/2 - m_t$ 
where $m_t$ is the top-quark mass \cite{Barbieri:2000vh}.  
In this case, after the production, the LSP 
squarks will hadronize by picking up $u$ or $d$ quark becoming 
charged or neutral fermionic mesons.  Since the charged one will be 
sufficiently long-lived to traverse the entire detector, it will be 
seen as highly ionizing tracks in future experiments.

\section{Conclusions}
\label{section:conc}

In this paper we have introduced a new electroweak symmetry breaking 
(EWSB) mechanism.  It has some similarities with the well-known 
supersymmetric radiative EWSB: the negative Higgs mass-squared arises 
from a one-loop diagram involving the large top-quark Yukawa coupling.  
However, our mechanism is inherently extra-dimensional: the Higgs 
mass-squared is determined by the compactification scale and is finite.

\begin{figure}
\begin{center} 
\begin{picture}(350,180)(-10,-25)
  \Line(60,-15)(60,125) \Line(270,15)(270,155)
  \Line(60,-15)(270,15) \Line(60,125)(270,155)
  \DashLine(125,20)(125,45){3} \DashLine(125,85)(125,110){3}
  \Text(120,20)[r]{$h$} \Text(120,110)[r]{$h$}
  \Vertex(125,45){2} \Vertex(125,85){2} \CArc(125,65)(20,0,360)
  \Text(100,65)[r]{$t$}
  \DashLine(195,30)(195,75){3} \DashLine(195,75)(195,120){3}
  \Text(190,30)[r]{$h$} \Text(190,120)[r]{$h$}
  \Vertex(195,75){2} \DashCArc(215,75)(20,0,360){3}
  \Text(240,75)[l]{$\tilde{t}$}
\end{picture}
\caption{Conventional supersymmetric radiative EWSB occurring in three
spatial dimensions.}
\label{fig:EWSB-1}
\end{center}
\end{figure}
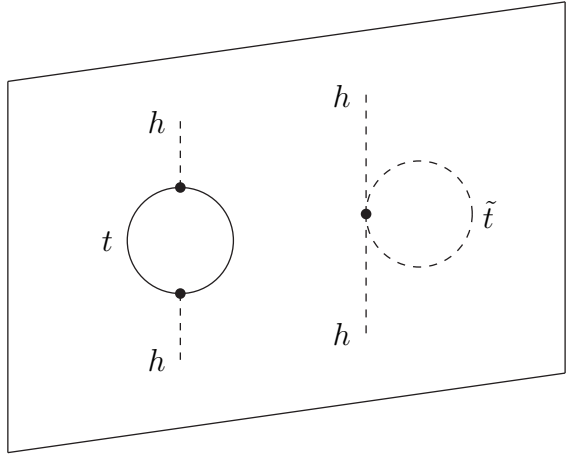
The conventional supersymmetric radiative EWSB is illustrated in 
Fig.~\ref{fig:EWSB-1}.  The particles, interactions and supersymmetry 
breaking are all located on the three brane.  While the quadratic 
divergence cancels between the two diagrams there is a residual 
logarithmic divergence, so that EWSB is being generated at all energy 
scales up to the scale at which the squark mass becomes soft.
This ranges from 100 TeV in some gauge mediated theories to the Planck 
scale in theories with gravity mediation of supersymmetry breaking.

\begin{figure}
  \centerline{
  \psfig{file=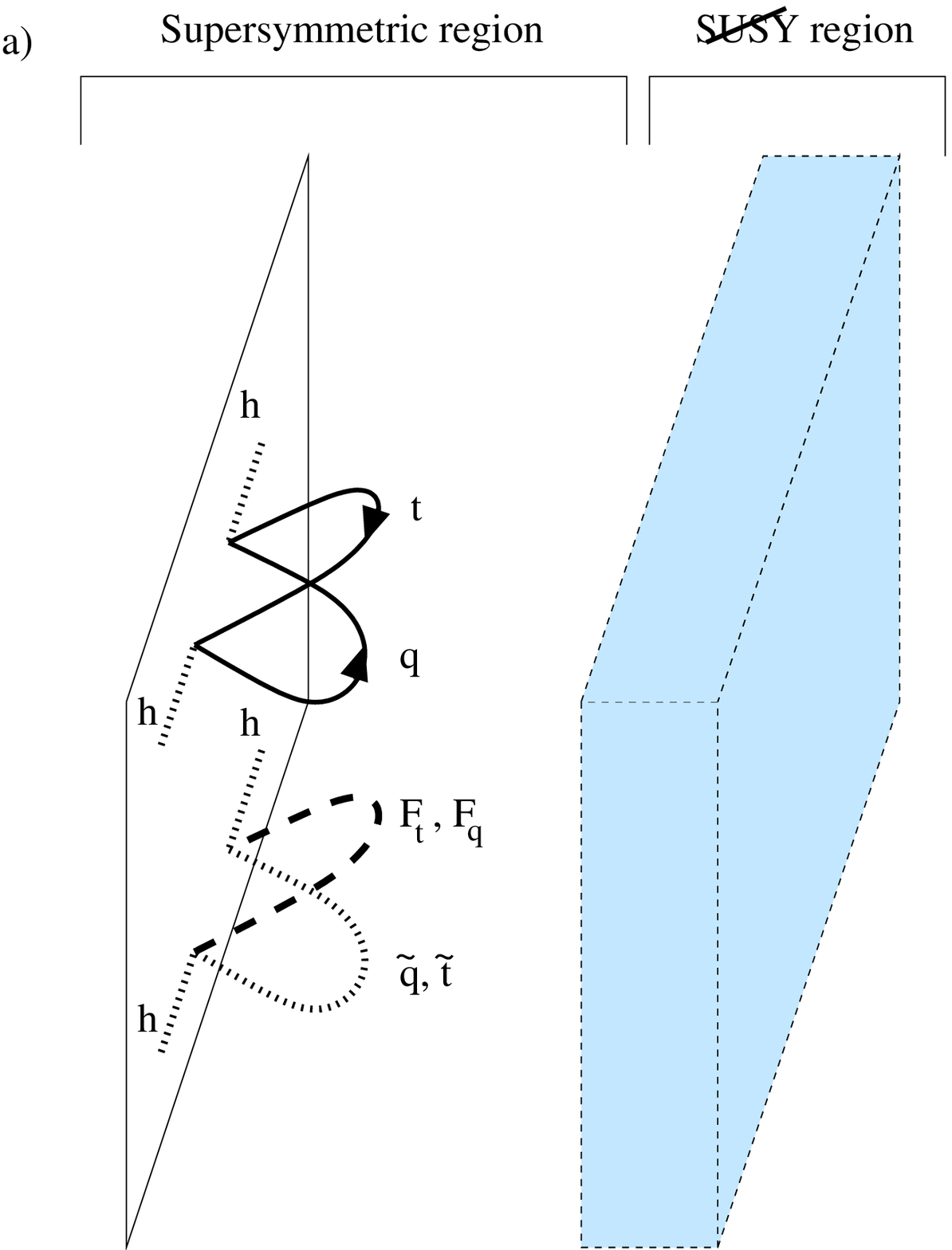,width=0.3\textwidth}\hskip 0.25in
  \psfig{file=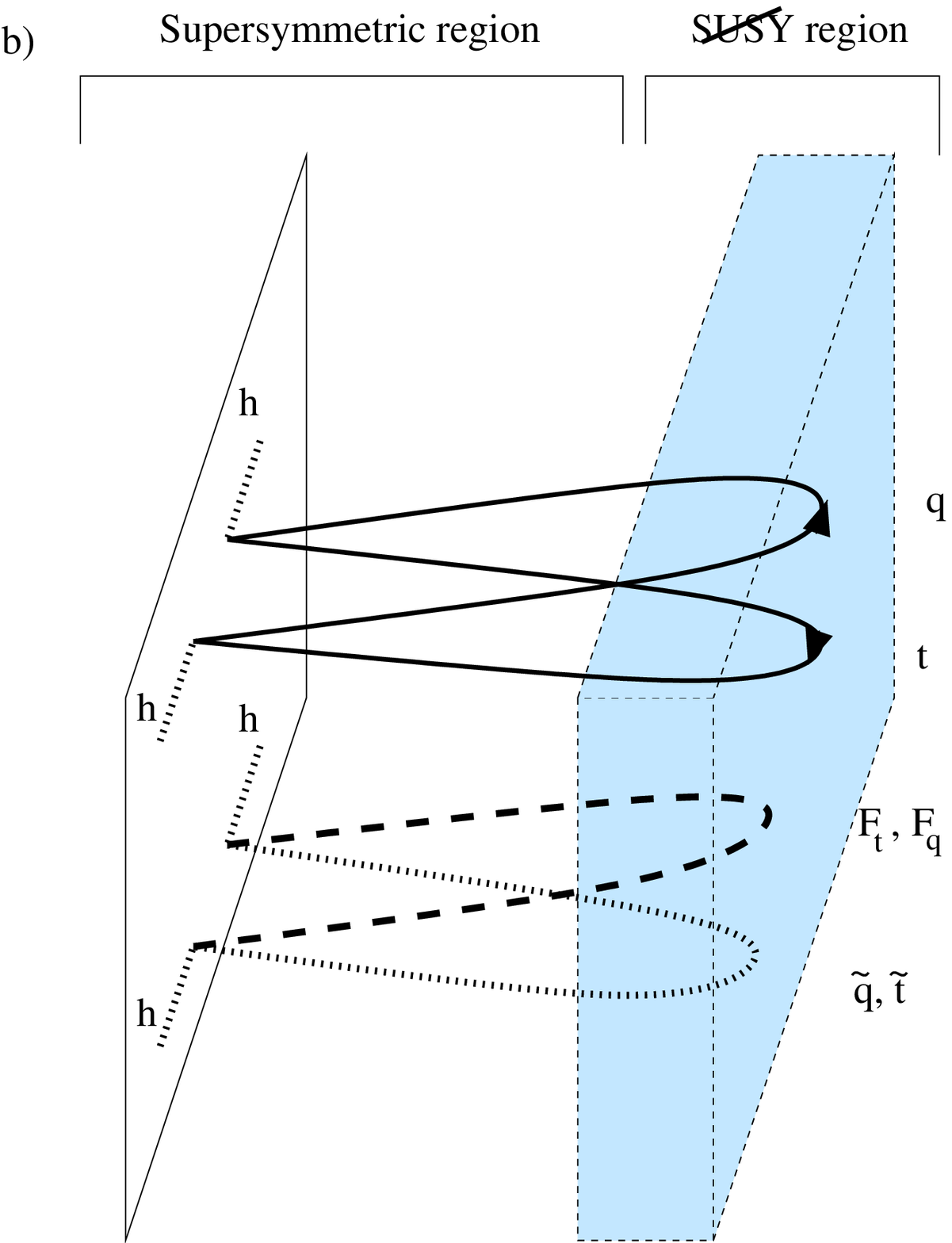,width=0.3\textwidth}}
\caption{Loops near the brane (a) do not sense supersymmetry breaking and
completely cancel.  Only once the fields propagate far from the Yukawa
brane (b) do they ``notice'' that supersymmetry is broken, for example by
boundary conditions or localized supersymmetry breaking.}
\label{fig:EWSB-2}
\end{figure}

The crucial feature of our mechanism is that the top quark propagates 
in a bulk of size $1/R \sim {\rm TeV}$, and the supersymmetry breaking 
is not located on the brane where the top-quark Yukawa coupling 
resides.  Thus the diagrams of Fig.~\ref{fig:EWSB-1}, where the internal 
top and stop particles are restricted to the three brane, are exactly 
supersymmetric and completely cancel.  A lack of cancellation only 
occurs when virtual particles propagate far into the bulk, as shown 
in Fig.~\ref{fig:EWSB-2}.

In fact, supersymmetry breaking is only significant if the virtual 
particle reaches a distance of order $R$ from the three brane.
If the virtual particle carries 4-momentum $k$, then propagation out 
to a distance $R$ is suppressed by $\exp(-k R)$; the virtual-momentum 
$k$ acts like a mass in the fifth dimension.  Hence the 4d momentum 
integral for the Higgs mass is exponentially damped for momenta above 
the compactification scale $1/R$.

Calculating the diagram of Fig.~\ref{fig:EWSB-2}, 
in momentum space for the usual four coordinates but in position space 
for the bulk, gives a Higgs mass-squared parameter:
\begin{equation}
  m_H^2 = -N_c f^2 \int\!\!dy_1\, dy_2 
  \, \delta(y_1) \, \delta(y_2) \int \!\!\frac{d^4 k}{(2\pi)^4}
  \frac{1}{2} \left( \tilde{G}_\psi^2 - \tilde{G}_\phi \tilde{G}_F \right).
\label{eq:general}
\end{equation}
The delta functions fix the Yukawa interactions on the brane at $y=0$, 
and the strength of the Yukawa coupling in 5d is $f = 2\pi R y_t$, 
as the Higgs is either a brane field or the zero mode of a bulk field.
The propagators of the top-quark chiral multiplet, 
$\tilde{G}_{\phi,\psi,F}$ for scalar, fermion and $F$ components, 
are here normalized such that in a non-compactified fifth dimension 
$\tilde{G}_{\phi,\psi,F} = 1$, giving an exact 
supersymmetric cancellation in Eq.~(\ref{eq:general}).  Since the delta
functions force the propagators to start and finish at $y=0$, one might 
guess that they do not probe whether the space is compactified.
This guess is incorrect: in a compactified space the propagation can 
wind $n$ times around the space (or reflect $n$ times), inducing 
$\tilde{G}_{\phi,\psi,F} = \tilde{G}_{\phi,\psi,F}(k R)$.
The form of the $k R$ dependence is determined by how 
supersymmetry is broken and how $\phi, \psi$ and $F$ feel this 
breaking.  In any theory in which there is no supersymmetry breaking 
locally near $y=0$, but rather has supersymmetry breaking on scales 
of order $\pi R$ from the Yukawa brane, then, for large $k$, 
$\tilde{G}_{\phi,\psi,F} = 1 + O(\exp(-k \pi R))$.  This ensures the 
momentum integral gives a finite $O(1/(\pi R)^4)$ result, so that 
$m_H^2 \simeq - N_c\, y_t^2/(\pi^4 R^2)$.  This general argument 
demonstrates that the physics of EWSB is strongly dominated by the 
scale $1/(\pi R)$ and is insensitive to how our 5d theory behaves 
when it becomes strongly coupled at larger energy scales.

In the models of sections \ref{section:localized} and 
\ref{section:Z2R-SS}, supersymmetric propagators have 
$\tilde{G} = \tanh[k \pi R]$, while those which feel supersymmetry
breaking have $\tilde{G} = \coth[k \pi R]$. In both models the quark
propagators are supersymmetric and the squark propagators feel
supersymmetry breaking. The numerical difference in $m_H^2$ arises
because in the Scherk-Schwarz theory the conjugate squarks (or $F$)
propagator feels supersymmetry breaking, while in the case of a
supersymmetry-breaking brane they do not.

Our new mechanism for radiative EWSB in the bulk has several important 
consequences for collider experiments. Since the higher dimensional
theory possesses more supersymmetries than in 4d theories, there are more
superpartners for particles which propagate in the bulk, which must
include $t,b$ quarks and the standard-model gauge particles, and might
include all matter. For example a top quark is accompanied by two
scalar superpartners $\tilde{t}$ and  $\tilde{t}^c$, as well as a
fermionic superpartner $t^c$. All these superpartners lie in the TeV
domain. Furthermore the superpartner masses are heavily influenced by
the geometry of the bulk: all matter superpartners which propagate in
the bulk, $\tilde{X}$, are highly degenerate, and similarly the conjugate
matter superpartners, $\tilde{X}^c$, are all degenerate. This
degeneracy implies that the flavor violation induced by squark or
slepton exchange is mild, and not problematic as frequently
encountered in 4d. 

There are KK resonances for all standard-model particles which
propagate in the bulk, and for all their superpartners. The mass
spectra of these towers are regular with mass splittings $1/R$. Two
simple examples of such spectra are shown in 
Figs.~\ref{fig:spectrum-1} and \ref{fig:spectrum-2}.

The Higgs sector of the theory is model dependent. For example the
Higgs doublets and their superpartners might be bulk fields or reside
on the brane where the Yukawa couplings are located. While a light
Higgs frequently occurs, it is also easy to violate the upper bound on
the lightest Higgs-boson mass of 4d supersymmetric theories. For
example, the Higgs boson may evade this bound by radiative
contributions to the effective potential of the Higgs from the KK
tower of top quarks and squarks. A model where this occurs is given in
section \ref{section:Higgs}, where the Higgs mass is correlated with the
compactification scale as shown in Fig.~\ref{fig:mudep}.

Two simple models discussed in this paper have unusual collider
phenomenology. In one case the LSP is a top squark. In collider
experiments this will lead to events with 1 or 2 highly ionizing charged  
tracks. In another case colored superpartners cascade decay to give
Higgs bosons and gravitinos: $\tilde{g} \rightarrow \bar{q} q
\tilde{h}$, followed by $\tilde{h} \rightarrow h \tilde{G}$, leading
to remarkable events with two Higgs bosons, missing transverse energy
and jets.

An important and generic consequence of our higher dimensional scheme
for breaking weak interactions is that the superpartners are typically
significantly heavier than in many 4d supersymmetric theories. In 4d
theories the Higgs mass-squared parameter and the mass-squared
parameters for squarks and sleptons occurs at the same order in
perturbation theory. Color factors or careful parameter choices can
push up the squark and slepton masses to some degree, but the
expectation is that the masses will be at the scale of the electroweak 
vacuum expectation value.
By contrast in our scheme matter superpartners acquire a tree-level
mass from propagation in the bulk, while the Higgs mass-squared
parameter is driven negative at the one-loop level. Superpartners which
propagate in the bulk we find to be typically a loop factor of
$\pi^2/2 \simeq 5$  heavier than the mass scale of the
radiatively generated Higgs mass. 
If all the superpartners propagate in the bulk, it is quite plausible
that none will be lighter than 1 TeV.

\section*{Acknowledgements}

We thank Riccardo Barbieri for useful conversations.
This work was supported by the Department of Energy under contract 
DE-AC03-76SF00098 and the National Science Foundation under 
contract PHY-95-14797.

\end{document}